\documentclass[aip,jmp,amsmath,amssymb,reprint]{revtex4-1}
\usepackage[latin9]{inputenc}
\usepackage{amsmath}
\usepackage{amssymb}
\usepackage{graphicx}

\makeatletter

\providecommand{\tabularnewline}{\\}

\usepackage{graphicx}
\usepackage{dcolumn}
\usepackage{bm}
\usepackage{subfigure}
\usepackage{color}
\usepackage{epstopdf}
\usepackage{nicefrac}

\makeatother

\begin{document}

\preprint{AIP/123-QED}

\title[Mass diffusion and liner material effect in a MagLIF fusion-like plasma]{Mass diffusion and liner material effect in a MagLIF fusion-like
plasma}

\author{F. Garc\'{i}a-Rubio}
\email{fernando.garcia.rubio@upm.es}

\affiliation{E.T.S.I. Aeronáutica y del Espacio, Universidad Politécnica de Madrid,
Madrid 28040, Spain.}

\author{J. Sanz}

\affiliation{E.T.S.I. Aeronáutica y del Espacio, Universidad Politécnica de Madrid,
Madrid 28040, Spain.}

\date{\today}
\begin{abstract}
In this paper, liner - fuel mass diffusion and the effect of the liner
material on mass ablation, energy and magnetic flux losses are studied
in a MagLIF fusion-like plasma. The analysis performed in {[}García-Rubio
and Sanz, Phys. Plasmas \textbf{24}, 072710 (2017){]} is extended
to liner materials of an arbitrary atomic number. The liner ablates
and penetrates into the hot spot, compressing and cooling down the
fuel. The magnetic flux in the fuel is lost by Nernst convection through
the ablated liner - fuel interface, called ablated border. Increasing
the liner atomic number leads to a reduction of both energy and magnetic
flux losses in the fuel for a small and moderate magnetization values.
Mass diffusion is confined within a thin layer at the ablated border.
Concentration gradient and baro-diffusion are the predominant mechanisms
leading to liner - fuel mixing. The width of the diffusion layer may
be comparable to the turbulent mixing layer resulting from the Rayleigh-Taylor
instability at the ablated border. An asymptotic analysis performed
for large liner atomic number $Z_{2}$ shows that mass ablation, energy
and magnetic flux losses and liner - fuel mass diffusion scale as
$1/\sqrt{Z_{2}}$. 
\end{abstract}
\maketitle

\section{Introduction\label{sec:Introduction}}

In the recently proposed magnetized liner inertial fusion (MagLIF)
scheme, a pulsed power machine drives the implosion of a conductive
cylindrical liner filled fuel that is magnetized and preheated\cite{slutz2010pulsed,gomez2014experimental}.
The advantages of magnetizing the fuel lie both on reducing heat losses\cite{landshoff1949transport}
and enhancing $\alpha$-particles energy deposition\cite{basko2000ignition}.
The liner is typically made of low atomic number metals such as lithium,
beryllium or aluminum. MagLIF concept has been scaled down in size
to OMEGA laser facility\cite{davies2017laser,barnak2017laser}, where
a laser drives the implosion of a parylene-Z plastic liner, less dense
than metal liners. 

A relevant feature of magnetized implosions is the long time that
the hot spot and the cold liner stay in contact. Understanding the
effect of fuel magnetization on heat and magnetic flux losses through
the hot spot - liner interface becomes essential. For this purpose,
the evolution of a hot magnetized plasma (hot spot) in contact with
a cold unmagnetized liner has recently been investigated in planar
geometry\cite{velikovich2015magnetic,garcia2017mass}. The problem
is studied in the low Mach number and high thermal to magnetic pressure
ratio $\beta$ limit, which implies isobaricity. In Ref. \onlinecite{garcia2017mass},
the liner is treated as a cold dense plasma made of the same material
as the fuel (deuterium), aiming to model the cryogenic fuel layer
added on the inner side of the liner in high-gain MagLIF configurations\cite{slutz2012high}.
The hot spot - liner interface represents an ablation front through
which the liner ablates, penetrates into the hot spot and cools it
down by thermal conduction. The interface separating the ablated liner
material and the fuel is referred to as ablated border. In Ref. \onlinecite{garcia2018magnetic},
this study was revisited including finite pressure ratio $\beta$
effects. 

\begin{figure}
\includegraphics[scale=0.35]{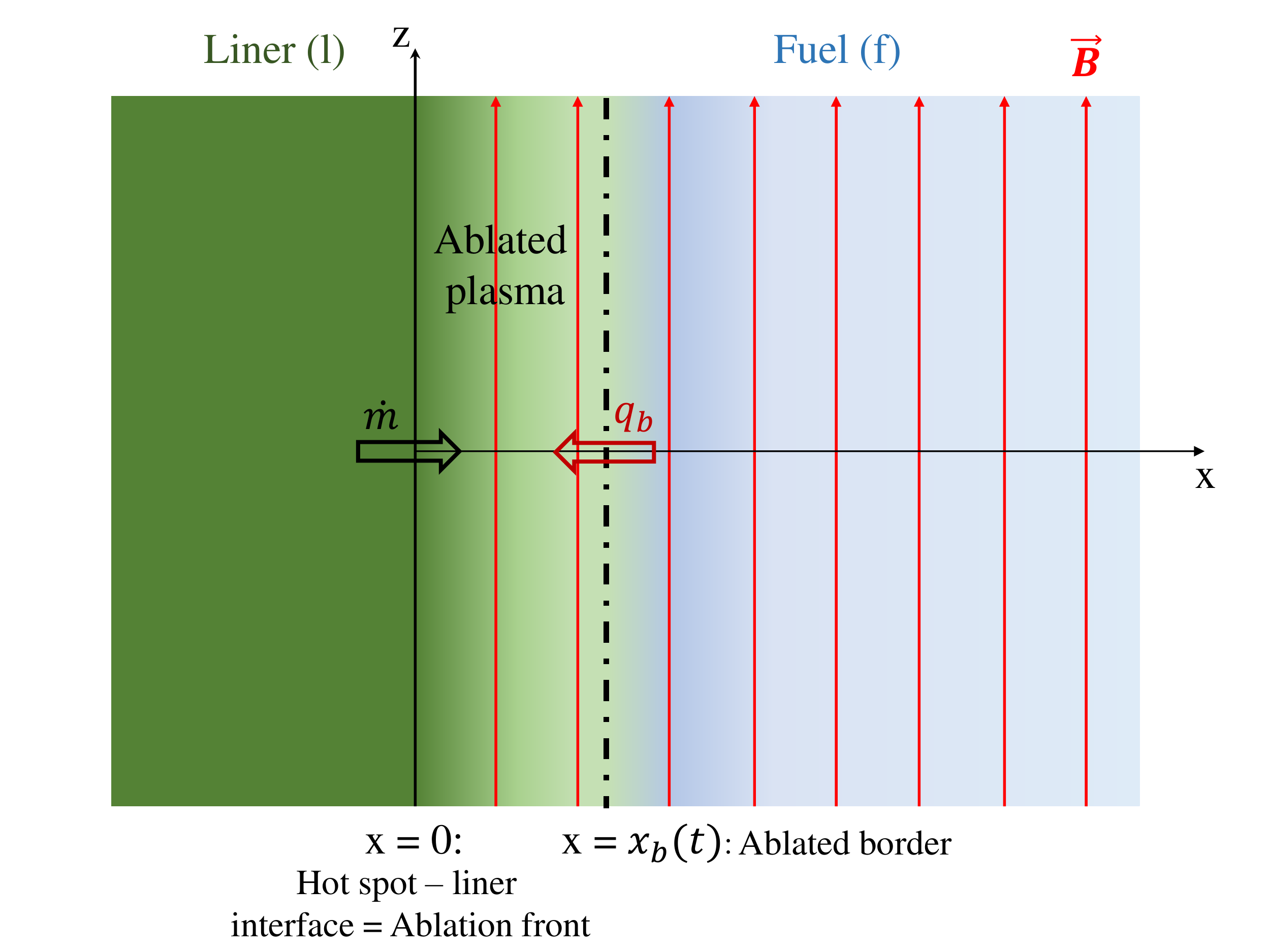}

\caption{Artistic scheme of the problem presented. The color gradation represents
ion density. Liner material (l) is shown in green, while fuel material
(f) is shown in blue. Arrows in red stand for magnetic field lines.
\label{fig:Artistic scheme}}
\end{figure}

In this paper, we extend the analysis performed in Ref. \onlinecite{garcia2017mass}
to liners made of an arbitrary material, see Fig. \ref{fig:Artistic scheme}.
The plasma ablated into the hot spot has therefore a different atomic
number $Z$ compared to the fuel. In the first part of this paper,
the two plasmas are treated as immiscible, and the ablated border
appears as a contact discontinuity where certain jump conditions must
be satisfied. In the second part of this paper, we let the two plasmas
diffuse and analyze liner and fuel mixing. 

Diffusion of ion species in multi-component plasmas has recently gained
attention as a potential mechanism explaining the observed yield anomalies
in inertial confinement fusion (ICF) implosions\cite{casey2012evidence,amendt2010barodiffusion}.
In plasmas, such diffusion can be driven by concentration, pressure,
electrostatic potential and temperature gradients. The latter three
are defined as baro-diffusion, electro-diffusion and thermo-diffusion,
respectively. Kagan and Tang\cite{kagan2012electro,kagan2014thermodynamic}
obtained the baro and electro-diffusion ratios $k_{p}$, $k_{e}$
without invoking a collisional model, proving that both ratios are
a thermodynamic quantity. In the authors' words, baro and electro-diffusion
are ``inextricably intertwined'', and while $k_{p}$ and $k_{e}$
depend on the choice of the thermodynamic system (including ions only
or the plasma as a whole), the overall diffusive flux stays the same
regardless of this choice\cite{kagan2014thermodynamic}. When the
thermodynamic system includes the ions only, the baro-diffusion ratio
$k_{p}$ turns out to be identical to its counterpart in a neutral
gas mix\cite{landau1959fluid}, and the electro-diffusion ratio $k_{e}$
depends on the charge-to-mass ratio of the ion species\cite{kagan2012electro}.
Thermo-diffusion, on its part, arises from the thermal force in the
friction drag between different ion species and between electrons
and ions. It depends on the nature of the collisions between particles.
Due to the long range of the Coulomb collisions, it plays a more important
role in plasmas compared to neutral gas mixtures. In addition, ion
thermo-diffusion reinforces baro-diffusion in plasmas, in contrast
to the neutral mixture case\cite{kagan2014thermodiffusion}.

In the work done by Simakov, Molvig and collaborators\cite{simakov2014electron,molvig2014classical,simakov2016hydrodynamicI,simakov2016hydrodynamicII},
the kinetic equations of multi-component plasmas are solved by a generalized
Chapman-Enskog expansion that assumes small Knudsen numbers for all
species. The transport terms are obtained, including ion species diffusion,
in the absence of magnetic field. In Ref. \onlinecite{molvig2014classical},
the electron and ion transport theories are thoroughly developed for
a two ion species plasma with disparate masses. The authors solve
the variational problem by expanding the density distribution in generalized
Laguerre polynomials of order 3/2, following the formalism of Helander
and Sigmar\cite{helander2005collisional}. The calculation is reduced
to two independent ``Spitzer'' problems for electrons and light
ions. The results for the electron transport theory are in agreement
with the results derived in Ref. \onlinecite{simakov2014electron},
where a generalization of Braginskii electron fluid description\cite{braginskii1965transport}
is performed for plasmas with multiple ion species, and therefore
used a different set of thermodynamic forces and fluxes that lead
to a different variational principle. In more recent publications\cite{simakov2016hydrodynamicI,simakov2016hydrodynamicII},
Simakov and Molvig developed the ion transport theory for an unmagnetized
collisional plasma with multiple ion species of arbitrary mass. The
fluid equations for multi-component plasmas are closed with the expressions
derived therein for individual ion species drift velocities, total
ion heat flux and viscosity. They followed a generalization of the
Braginskii ion fluid description, which agrees with the results previously
obtained for two ions species with disparate masses in Ref. \onlinecite{molvig2014classical}. 

We apply the theory developed in Refs. \onlinecite{simakov2016hydrodynamicI}
and \onlinecite{simakov2016hydrodynamicII} in the second part of
this paper to study mass diffusion at the ablated border. Since the
transport terms have only been derived for an unmagnetized plasma,
we isolate the hydrodynamic problem and disregard the magnetic field
evolution in this part. This paper is therefore organized as follows.
In Sec. \ref{sec:governing-equations}, the problem is presented and
the governing equations are discussed not taking into account mass
diffusion. They can be reduced to a system of two partial differential
equations for temperature and magnetic field in which the solution
presents a self-similar structure. In Sec. \ref{sec:analysis-and-results},
the results are discussed and the effect of the liner material on
the thermal and magnetic flux losses is studied. In Sec. \ref{sec:governing-equations-with-mass-transport},
the problem is formulated again taking into account mass diffusion.
The governing equations can be reduced to a system of two independent
equations for temperature and fuel concentration, whose solution also
presents a self-similar nature. In Sec. \ref{sec:results-with-mass-transport},
the results with mass diffusion are presented and in Sec. \ref{sec:conclusions},
conclusions are drawn.

\section{governing equations without mass diffusion\label{sec:governing-equations}}

We consider at $t=0$ a hot plasma medium at rest with a uniform temperature
$T_{0}$, ion particle density $n_{0}$ and thermal pressure $p_{0}$
occupying the semi-infinite space $x>0$. It is initially magnetized
with a uniform magnetic field $\vec{B}_{0}=B_{0}\vec{e}_{z}$. The
hot plasma is in contact at $x=0$ with a cold denser unmagnetized
plasma $(T\ll T_{0},|\vec{B}|\ll B_{0})$ that extends towards $x<0$,
and they are in mechanical equilibrium (same pressure). Both plasmas
are fully ionized and made of a different material, that is, different
atomic number $Z$. They crudely represent the fuel, $Z=Z_{1}$, and
the liner, $Z=Z_{2}$, respectively. Hereinafter, we consider that
the fuel is made of deuterium, $Z_{1}=1$, and we take $Z_{2}$ as
a free parameter. We let the system evolve for $t>0$. As a consequence
of thermal conduction, the liner material ablates, compresses the
fuel and cools it down, while the magnetic field is subjected to convection
and diffusion. The region $x>0$, which stands for the hot spot, is
therefore formed of two plasmas of different atomic number $Z$: ablated
liner material and fuel, with the ablated border being the interface
between them, see Fig. \ref{fig:Artistic scheme}. The hot spot -
liner interface, $x=0$, represents consequently an ablation front.
The liner (cold dense plasma), stays at rest and unmagnetized since
its thermal conductivity, $\chi\sim T^{5/2}$, is low and the magnetic
diffusivity, $D_{m}\sim T^{-3/2}$, is high (it can be considered
as a magnetic isolant); neither heat flux nor electrical currents
can take place in it. 

We therefore focus on the dynamics of the plasmas in the hot spot,
$x>0$. As typically occurs in MagLIF implosions\cite{velikovich2015magnetic},
we assume subsonic motion (low Mach number), and large thermal to
magnetic pressure ratio: $\beta=8\pi p_{0}/B_{0}^{2}\gg1$. Due to
the geometry of the problem, the only independent variables are the
streamwise direction $x$ and time $t$, the plasma ion velocity is
one-dimensional $\vec{v}=v\left(x,t\right)\vec{e}_{x}$, and the magnetic
field stays perpendicular to the motion $\vec{B}=B\left(x,t\right)\vec{e}_{z}$.
Quasi-neutrality, same ion and electron temperatures and ideal gas
hypotheses are assumed, and we take the adiabatic index $\gamma=5/3$.
The evolution of the ion particle density $n$, thermal pressure $p=p_{e}+p_{i}$,
temperature $T$, ion velocity $v$ and magnetic field $B$ is governed
by the ion continuity, plasma momentum and energy conservation equations
neglecting plasma viscosity, together with Faraday's law (induction
equation) and the equation of state, which for a small Mach number
and large $\beta$ limit read

\begin{equation}
\dfrac{\partial n}{\partial t}+\dfrac{\partial}{\partial x}\left(nv\right)=0,\label{eq:ion continuity}
\end{equation}

\begin{equation}
\dfrac{\partial p}{\partial x}=0,\label{eq:total momentum}
\end{equation}

\begin{equation}
\dfrac{1}{\gamma-1}\dfrac{\partial p}{\partial t}+\dfrac{\partial}{\partial x}\left(\dfrac{\gamma}{\gamma-1}pv\right)=\dfrac{\partial}{\partial x}\left(\underbrace{\chi_{\perp}\dfrac{\partial T}{\partial x}}_{\text{Cond.}}\right),\label{eq:total energy}
\end{equation}

\begin{equation}
\dfrac{\partial B}{\partial t}+\dfrac{\partial}{\partial x}\left(vB\right)=\dfrac{\partial}{\partial x}\left(\underbrace{D_{m\perp}\dfrac{\partial B}{\partial x}}_{\text{Joule}}+\underbrace{\dfrac{c\beta_{\wedge}^{uT}}{Zen}\dfrac{\partial T}{\partial x}}_{\text{Nernst}}\right),\label{eq:induction}
\end{equation}

\begin{equation}
p=(1+Z)nT.\label{eq:general equation of state}
\end{equation}
We use Braginskii's expressions and notations for the transport coefficients\cite{braginskii1965transport},
with a constant value equal to $7$ for the Coulomb logarithm $\lambda$.
The coefficient $\chi_{\perp}$ stands for electron plus ion conductivities,
$\beta_{\wedge}^{uT}$ refers to the transport coefficient for the
Nernst effect and $D_{m\perp}=\alpha_{\perp}c^{2}/4\pi e^{2}Z^{2}n^{2}$
is the magnetic diffusion coefficient appearing in the Joule dissipation.
The momentum equation \eqref{eq:total momentum} is reduced to isobaricity:
thermal pressure is constant and given by the initial conditions,
$p=p_{0}$. Plasma energy equation \eqref{eq:total energy} is thus
reduced to the balance between enthalpy convection and thermal conduction.
It can be integrated once, yielding an explicit relation for the ion
velocity

\begin{equation}
v=\dfrac{\gamma-1}{\gamma p_{0}}\chi_{\perp}\dfrac{\partial T}{\partial x}.\label{eq:velocity}
\end{equation}
Plasma velocity is therefore proportional to the thermal conduction
heat flux $q=-\chi_{\perp}\partial T/\partial x$. As discussed in
Ref. \onlinecite{garcia2017mass}, in the integration of the equation
of energy one should add a constant of integration $v_{\infty}$,
which happens to be zero if the liner is modeled as a cold dense plasma.
We impose zero heat flux at the liner interface since the heat conduction
losses are recycled back into the hot spot via the ablated material\cite{betti2001hot,sanz2005analytical}. 

The position of the ablated border, denoted by $x_{b}(t)$, has to
be determined self-consistently. It is determined by applying the
fluid surface condition\cite{garcia2017mass}, 
\begin{equation}
\dfrac{\text{d}x_{b}}{\text{d}t}=v\left(x_{b},t\right).\label{eq:xb}
\end{equation}
In this part of the paper, we consider that both plasmas are immiscible
(no atomic mass diffusion), hence $x_{b}(t)$ represents also a contact
discontinuity. Consequently, the atomic number $Z$ takes the value
$Z=Z_{2}$ (liner material) at $x<x_{b}(t)$ and $Z=Z_{1}$ (fuel)
at $x>x_{b}(t)$. 

The system of governing equations \eqref{eq:ion continuity}, \eqref{eq:induction}
and \eqref{eq:velocity} has to be completed with the appropriate
boundary conditions. We impose that far from the liner, $x\rightarrow\infty$,
the plasma recovers the initial fuel state $n=n_{0}$, $T=T_{0}$
and $B=B_{0}$. At the interface with the liner, $x=0$, we require
that the plasma temperature and magnetic field be equal to the temperature
and magnetic field values at the liner, that is $T/T_{0}\rightarrow0$
and $B/B_{0}\rightarrow0$. 

In addition to the boundary conditions, we demand that temperature,
magnetic field and ion velocity be continuous at the ablated border.
The latter condition implies continuity of the heat flux $q$. According
to Eq. \eqref{eq:general equation of state}, the ion particle density
is discontinuous, and it satisfies the relation
\begin{equation}
\left(Z_{2}+1\right)n\left(x_{b}^{-},t\right)=\left(Z_{1}+1\right)n\left(x_{b}^{+},t\right).\label{eq:density jump}
\end{equation}
We also introduce the mass density as $\rho=m_{i}n$, with $m_{i}$
being the ion mass. For simplicity, we consider that in both fuel
and liner materials, the mass number doubles the atomic number, and
we take the ion mass $m_{i}$ to be related to the proton mass $m_{p}$
through $m_{i}=2Zm_{p}$. The former jump condition turns into
\begin{equation}
\dfrac{Z_{2}+1}{Z_{2}}\rho\left(x_{b}^{-},t\right)=\dfrac{Z_{1}+1}{Z_{1}}\rho\left(x_{b}^{+},t\right).\label{eq:mass density jump}
\end{equation}
Notice that for $Z_{1}=1$, $n\left(x_{b}^{-},t\right)\leq n\left(x_{b}^{+},t\right)$
but $\rho\left(x_{b}^{-},t\right)\geq\rho\left(x_{b}^{+},t\right)$.

Finally, the induction equation \eqref{eq:induction} forces the sum
of the Joule dissipation plus the Nernst term $D_{m\perp}\partial B/\partial x+\left(c\beta_{\wedge}^{uT}/Zen\right)\partial T/\partial x$
to be continuous through the ablated border.

\subsection{Normalization and self-similarity\label{subsec:Normalization-and-self-similarit}}

\begin{figure}
\includegraphics[scale=0.25]{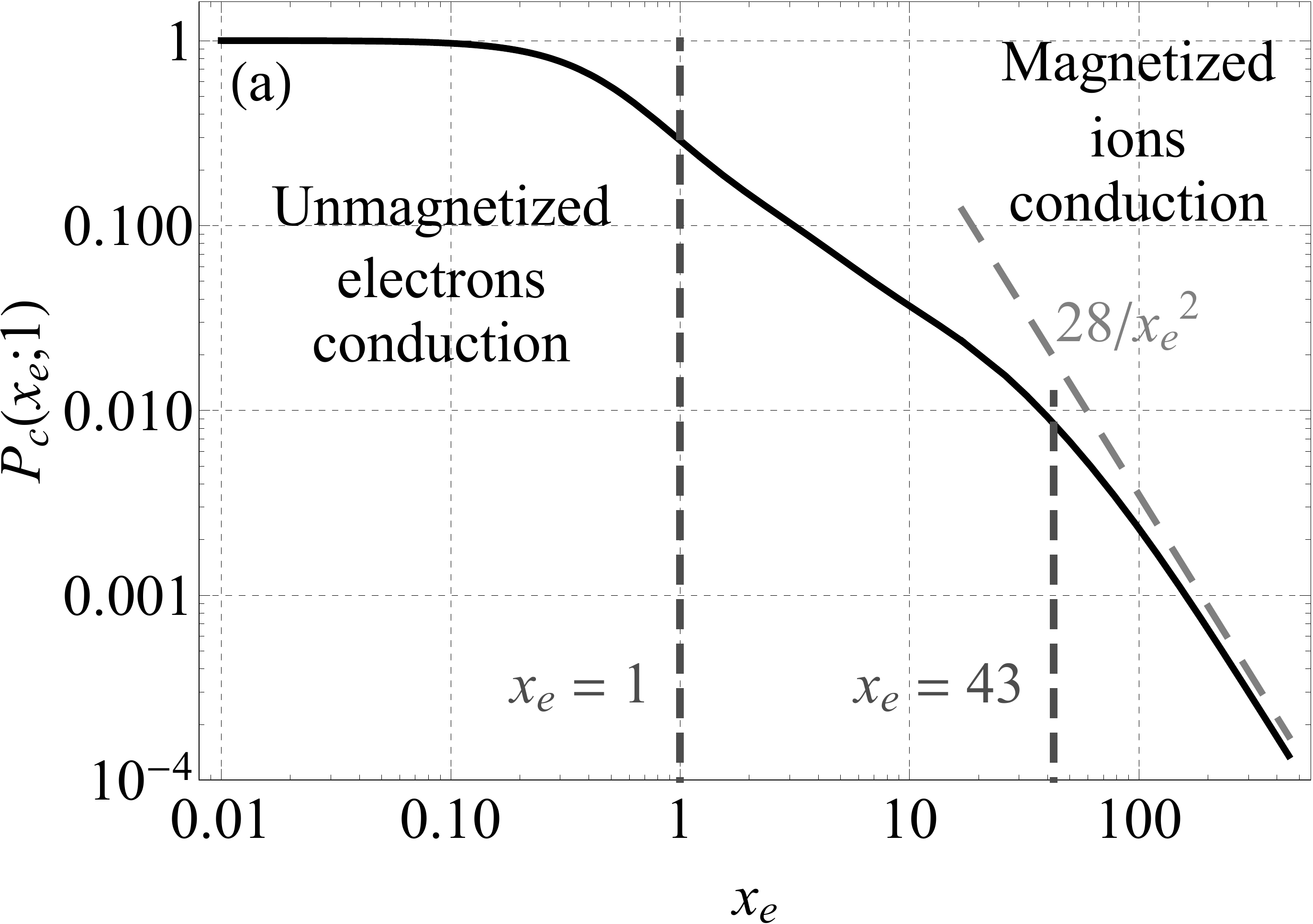}

\includegraphics[scale=0.25]{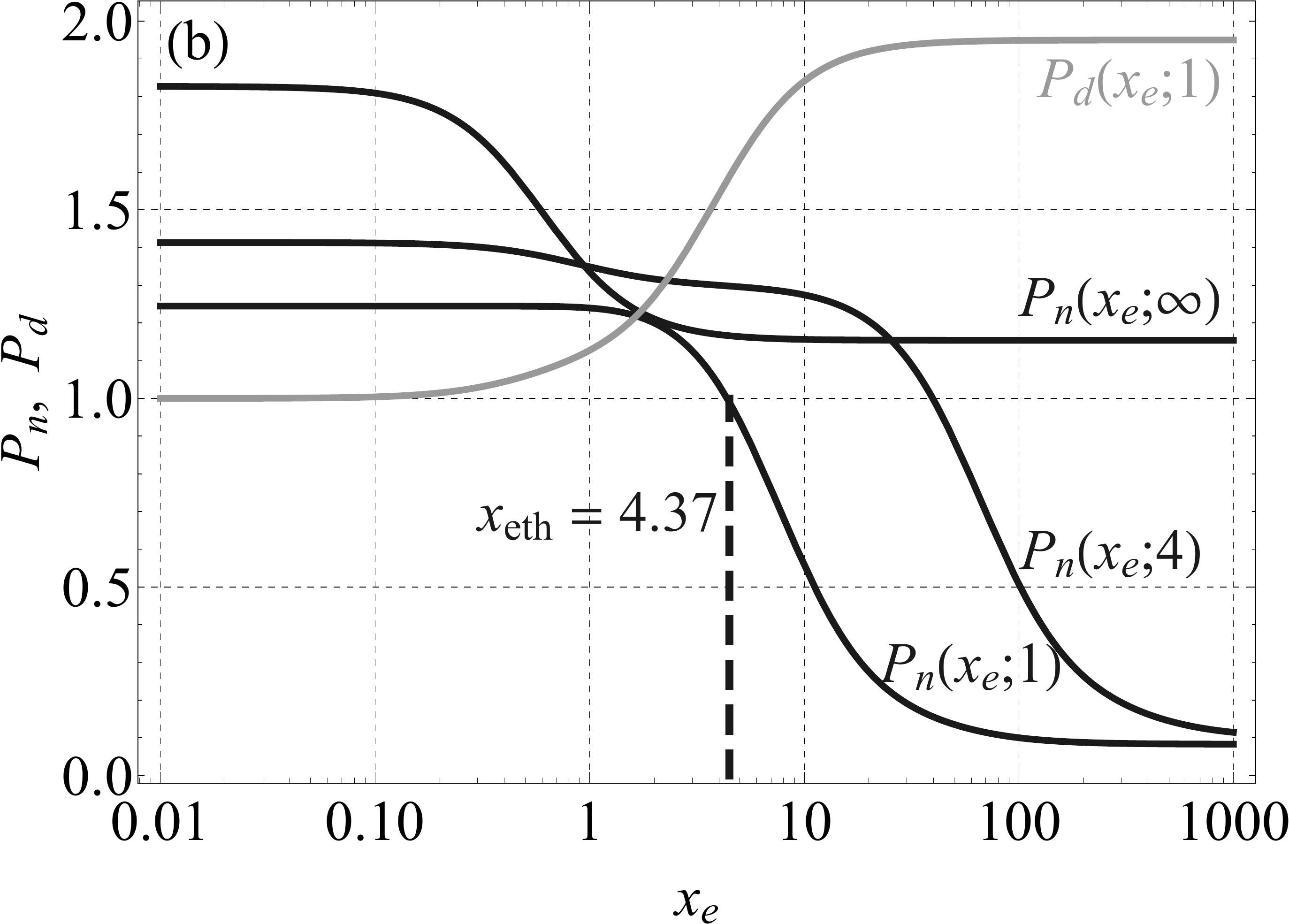}

\caption{\label{fig:Transport polynomials}Effect of magnetization on the transport
coefficients: thermal conduction $\mathcal{P}_{c}\left(x_{e};Z\right)$,
Nernst term $\mathcal{P}_{n}\left(x_{e};Z\right)$ and Joule dissipation
$\mathcal{P}_{d}\left(x_{e};Z\right)$.}
\end{figure}

The ion particle density, temperature and magnetic field in the hot
spot are normalized with their initial value in the fuel: $\sigma=n/n_{0}$,
$\theta=T/T_{0}$ and $\phi=B/B_{0}$. Ion density is related to temperature
through the equation of state \eqref{eq:general equation of state},
yielding $\sigma=\left(Z_{1}+1\right)/\left(Z+1\right)\theta$. The
ion mass density is likewise normalized $\varrho=\rho/\rho_{0}$,
with $\rho_{0}=2Z_{1}m_{p}n_{0}$. The thermal conductivity coefficient
is normalized with the unmagnetized electron conductivity, $\chi_{\perp}=\bar{K}T_{0}^{5/2}\theta^{5/2}\mathcal{P}_{c}\left(x_{e};Z\right)$,
where 
\begin{equation}
\bar{K}\equiv\dfrac{Zn\tau_{e}\gamma_{0}}{m_{e}T^{3/2}}=\dfrac{3\gamma_{0}}{4\sqrt{2\pi m_{e}}e^{4}\lambda Z}\label{eq:K no diff}
\end{equation}
is the constant factor in Spitzer conductivity, and $\tau_{e},m_{e}$
are the electron collision time and mass, respectively. The thermoelectric
transport coefficient is proportional to the thermal conductivity
coefficient $c\beta_{\wedge}^{uT}/Zen=\left[\left(\gamma-1\right)\chi_{\perp}/\gamma p_{0}\right]B\mathcal{P}_{n}\left(x_{e};Z\right)$,
while the magnetic diffusion coefficient is written as $D_{m\perp}=\bar{D}T_{0}^{-3/2}\theta^{-3/2}\mathcal{P}_{d}\left(x_{e};Z\right)$,
with 
\begin{equation}
\bar{D}\equiv\dfrac{c^{2}m_{e}T^{3/2}\alpha_{0}}{4\pi e^{2}Zn\tau_{e}}=\dfrac{c^{2}\sqrt{2m_{e}}\lambda e^{2}\alpha_{0}Z}{3\sqrt{\pi}}\label{eq:D no diff}
\end{equation}
being a diffusivity constant. Notice that the conductivity and diffusivity
constants are functions of the atomic number: $\bar{K}\left(Z\right),\bar{D}\left(Z\right)$.
The terms $\mathcal{P}_{c}$, $\mathcal{P}_{n}$ and $\mathcal{P}_{d}$
account for the effect of magnetization on the transport coefficients,
see Fig. \ref{fig:Transport polynomials}. They are rational functions
of the electron Hall parameter (electron cyclotron frequency times
the electron collision time) $x_{e}=\omega_{e}\tau_{e}=\left(eB/m_{e}c\right)\tau_{e}$
and the atomic number $Z$, and read 
\[
\mathcal{P}_{c}\left(x_{e};Z\right)=\dfrac{\gamma_{1}^{\prime}x_{e}^{2}+\gamma_{0}^{\prime}}{\gamma_{0}\Delta_{e}}+\dfrac{1}{Z^{3}}\sqrt{\dfrac{2m_{e}}{m_{i}}}\dfrac{2x_{i}^{2}+2.645}{\gamma_{0}\Delta_{i}},
\]

\begin{multline*}
\mathcal{P}_{n}\left(x_{e};Z\right)=\dfrac{Z+1}{Z}\dfrac{\gamma}{\gamma-1}\times\\
\dfrac{\beta_{1}^{\prime\prime}x_{e}^{2}+\beta_{0}^{\prime\prime}}{\gamma_{1}^{\prime}x_{e}^{2}+\gamma_{0}^{\prime}+\dfrac{1}{Z^{3}}\sqrt{\dfrac{2m_{e}}{m_{i}}}\dfrac{\Delta_{e}}{\Delta_{i}}\left(2x_{i}^{2}+2.645\right)},
\end{multline*}
\[
\mathcal{P}_{d}\left(x_{e};Z\right)=\dfrac{1-\dfrac{\alpha_{1}^{\prime}x_{e}^{2}+\alpha_{0}^{\prime}}{\Delta_{e}}}{\alpha_{0}},
\]
with $\Delta_{e}=x_{e}^{4}+\delta_{1}x_{e}^{2}+\delta_{0}$ and $\Delta_{i}=x_{i}^{4}+2.70x_{i}^{2}+0.677$,
and $x_{i}=\omega_{i}\tau_{i}=x_{e}\sqrt{2m_{e}/m_{i}}/Z$ standing
for the ion magnetization. Finally, the coefficients $\gamma_{0}$,
$\alpha_{0}$, $\gamma_{1}^{\prime}$ along with others are functions
of the atomic number and are given in Braginskii\cite{braginskii1965transport}
for $Z=1,2,3,4$ and $Z\rightarrow\infty$.

In an unmagnetized plasma, $x_{e}\ll1$, thermal conduction is mainly
due to electrons because of their small mass and $\mathcal{P}_{c}\approx1$.
However, the electrons get magnetized for lower magnetic field intensities
compared to the ions ($x_{e}=1$ with respect to $x_{i}=1\Rightarrow x_{e}=Z\sqrt{m_{i}/2m_{e}}\approx43Z^{3/2}$),
and when electron conduction is suppressed due to magnetization, the
ions carry the heat transport. For large magnetization values, $x_{e}>Z\sqrt{m_{i}/2m_{e}}$,
the ions also get magnetized and the thermal conductivity decreases
asymptotically as $\mathcal{P}_{c}\sim x_{e}^{-2}$.

The electron Hall parameter anywhere in the hot spot, $x_{e}$, can
be written as a function of its initial value in the fuel, the atomic
number $Z$ and the dimensionless temperature and magnetic field profiles
\begin{equation}
x_{e}=x_{e0}\dfrac{Z+1}{Z^{2}}\dfrac{Z_{1}^{2}}{Z_{1}+1}\phi\theta^{5/2}.\label{eq:electron Hall parameter}
\end{equation}

We define the magnetic Lewis number $\text{Le}_{\text{m}}$ as the
ratio between the thermal and magnetic diffusivities in the fuel,
defined, on their part, as 
\begin{equation}
\kappa_{0}\equiv\dfrac{\left(\gamma-1\right)\bar{K}\left(Z_{1}\right)T_{0}^{7/2}}{\gamma p_{0}}\approx\dfrac{40,500}{\lambda}\dfrac{T_{0}^{5/2}}{\rho_{0}}\text{ cm}^{2}/\text{sec}\label{eq:k0 no diff}
\end{equation}
 and 
\begin{equation}
\nu_{m0}\equiv\bar{D}\left(Z_{1}\right)T_{0}^{-3/2}\approx\dfrac{13.33\lambda}{T_{0}^{3/2}}\text{ cm}^{2}/\text{sec},\label{eq:num0}
\end{equation}
respectively. It reads
\begin{equation}
\text{Le}_{\text{m}}\equiv\dfrac{\kappa_{0}}{\nu_{m0}}\approx\dfrac{3038}{\lambda^{2}}\dfrac{T_{0}^{4}}{\rho_{0}}.\label{eq:Lewis number-1}
\end{equation}
In the practical formulas here and below, the temperature is expressed
in keV and the mass density in g/cc. The magnetic Lewis number can
also be expressed in terms of the pressure ratio $\beta$ and the
initial electron Hall parameter $x_{e0}$,
\begin{equation}
\text{Le}_{\text{m}}\equiv\dfrac{\kappa_{0}}{\nu_{m0}}=\dfrac{\gamma-1}{2\gamma}\dfrac{Z_{1}^{2}}{\left(Z_{1}+1\right)^{2}}\dfrac{\gamma_{0}\left(Z_{1}\right)}{\alpha_{0}\left(Z_{1}\right)}\beta x_{e0}^{2}\approx0.31\text{\ensuremath{\beta}}x_{e0}^{2}.\label{eq:Lewis number}
\end{equation}

The initial value problem proposed lacks a characteristic length and
velocity, and the ablated border position $x_{b}(t)$ is an eigenvalue
of the problem. Therefore, the solution to \eqref{eq:ion continuity},
\eqref{eq:induction} and \eqref{eq:velocity} is sought under the
form of a self-similar diffusive wave\cite{zel2002physics}, as done
in Refs. \onlinecite{garcia2017mass,velikovich2015magnetic}. We introduce,
consequently, the independent self-similar variable $\eta\geq0$

\begin{equation}
\eta=\dfrac{x}{\sqrt{\kappa_{0}t}}.\label{eq:slef-similar variable}
\end{equation}
Consistently, the plasma ion velocity is scaled self-similarly

\begin{equation}
V\left(\eta\right)=2\sqrt{\dfrac{t}{\kappa_{0}}}v(x,t),\label{eq:velocity ansatz}
\end{equation}
which, by the use of the equation of energy \eqref{eq:velocity},
can be linked to temperature and magnetic field profiles as

\begin{equation}
V=2\bar{K}_{R}\mathcal{P}_{c}\theta^{5/2}\dfrac{\text{d}\theta}{\text{d}\eta}.\label{eq:velocity selfsimilar}
\end{equation}
Here, $\bar{K}_{R}$ stands for the Spitzer conductivity constant
normalized with its value in the fuel, 
\begin{equation}
\bar{K}_{R}\left(Z\right)\equiv\dfrac{\bar{K}\left(Z\right)}{\bar{K}\left(Z_{1}\right)}=\dfrac{\gamma_{0}\left(Z\right)Z_{1}}{\gamma_{0}\left(Z_{1}\right)Z}.\label{eq:KR}
\end{equation}

The ion continuity equation \eqref{eq:ion continuity} is rewritten
as 
\begin{equation}
\left(V-\eta\right)\dfrac{\text{d}\sigma}{\text{d}\eta}+\sigma\dfrac{\text{d}V}{\text{d}\eta}=0,\label{eq:ion continuity self-similar}
\end{equation}
which, by the use of the equation of state and inserting the expression
for $V$, becomes 
\begin{equation}
\eta\dfrac{\text{d}\theta}{\text{d}\eta}+2\theta^{2}\dfrac{\text{d}}{\text{d}\eta}\left(\bar{K}_{R}\mathcal{P}_{c}\theta^{3/2}\dfrac{\text{d}\theta}{\text{d}\eta}\right)=0.\label{eq:ion continuity in T self-similar}
\end{equation}

The induction equation \eqref{eq:induction} likewise normalized gives:

\begin{multline}
-\eta\dfrac{\text{d}\phi}{\text{d}\eta}+2\dfrac{\text{d}}{\text{d}\eta}\left[\bar{K}_{R}\mathcal{P}_{c}\left(1-\mathcal{P}_{n}\right)\theta^{5/2}\dfrac{\text{d}\theta}{\text{d}\eta}\phi\right]\\
=\dfrac{2}{\text{Le}_{\text{m}}}\dfrac{\text{d}}{\text{d}\eta}\left(\bar{D}_{R}\dfrac{\mathcal{P}_{d}}{\theta^{3/2}}\dfrac{\text{d}\phi}{\text{d}\eta}\right).\label{eq:induction self-similar}
\end{multline}
Similarly, $\bar{D}_{R}$ is the diffusivity constant normalized with
its value in the fuel, 
\begin{equation}
\bar{D}_{R}\left(Z\right)\equiv\dfrac{\bar{D}\left(Z\right)}{\bar{D}\left(Z_{1}\right)}=\dfrac{\alpha_{0}\left(Z\right)Z}{\alpha_{0}\left(Z_{1}\right)Z_{1}}.\label{eq:DR}
\end{equation}
The conductivity ratio $\bar{K}_{R}\left(Z\right)$ is a decreasing
function of $Z$, while $\bar{D}_{R}\left(Z\right)$ is an increasing
function of $Z$. 

Two opposite effects take part in the convection term in Eq. \eqref{eq:induction self-similar}:
magnetic field convection by the plasma motion, in the opposite direction
to the conduction heat flux, and convection due to the Nernst term,
in the same direction. The balance between both effects is given by
the term $1-\mathcal{P}_{n}$. There is a threshold magnetization
value $x_{e\text{th}}$ below which $\mathcal{P}_{n}>1$, the Nernst
term is predominant and the magnetic field is convected towards the
liner. In the regions where $x_{e}>x_{e\text{th}},$ the Nernst terms
is less effective and the magnetic field is convected by the plasma
towards the center of the hot spot. The value of $x_{e\text{th}}$
increases with $Z$, see Fig. \ref{fig:Transport polynomials}(b).
It takes the value $x_{e\text{th}}=4.37$ for $Z=1$, while $\mathcal{P}_{n}$
is always greater than 1 for $Z\rightarrow\infty$.

The problem is therefore reduced to solving the equations \eqref{eq:ion continuity in T self-similar}
and \eqref{eq:induction self-similar}, which form a system of two
ordinary differential equations of fourth order for the normalized
temperature and magnetic field profiles $\theta\left(\eta\right)$,
$\phi\left(\eta\right)$. They are coupled through the dependence
of the thermal conductivity on the magnetization and need to be completed
with the previously established boundary conditions: 
\begin{equation}
\theta\left(0\right)=0,\quad\theta\left(\eta\rightarrow\infty\right)=1,\label{eq:BC for T}
\end{equation}
\begin{equation}
\phi\left(0\right)=0,\quad\phi\left(\eta\rightarrow\infty\right)=1.\label{eq:BC for B}
\end{equation}

The normalized position of the ablated border, $\eta_{b}=x_{b}/\sqrt{\kappa_{0}t}$,
has to be solved self-consistently. It is determined by Eq. \eqref{eq:xb},
which transforms into 
\begin{equation}
\eta_{b}=V\left(\eta_{b}\right).\label{eq:eta b}
\end{equation}
The eigenvalue $\eta_{b}$ is thus proportional to the heat flux at
the ablated border. Notice that, according to Eq. \eqref{eq:ion continuity self-similar},
the ion velocity has a stationary point there. The electron Hall parameter,
Eq. \eqref{eq:electron Hall parameter}, is discontinuous, and its
value to the right $x_{e}^{+}=x_{e}\left(\eta_{b}^{+}\right)$ is
larger than its value to the left $x_{e}^{-}=x_{e}\left(\eta_{b}^{-}\right)$,
related by
\begin{equation}
x_{e}^{+}=x_{e}^{-}\dfrac{Z_{2}^{2}}{Z_{2}+1}\dfrac{Z_{1}+1}{Z_{1}^{2}}.\label{eq:electron Hall jump}
\end{equation}
The continuity of the heat flux and the Joule plus Nernst terms establishes
the following conditions for the derivatives of the temperature and
magnetic field
\begin{equation}
\left.\dfrac{\text{d}\theta}{\text{d}\eta}\right|_{\eta_{b}^{+}}=\bar{K}_{R}\left(Z_{2}\right)\dfrac{\mathcal{P}_{c}^{-}}{\mathcal{P}_{c}^{+}}\left.\dfrac{\text{d}\theta}{\text{d}\eta}\right|_{\eta_{b}^{-}},\label{eq:temperature dev jump}
\end{equation}
\begin{multline}
\left.\dfrac{\text{d}\phi}{\text{d}\eta}\right|_{\eta_{b}^{+}}=\\
\text{Le}_{\text{m}}\dfrac{\theta_{b}^{4}}{\mathcal{P}_{d}^{+}}\bar{K}_{R}\left(Z_{2}\right)\mathcal{P}_{c}^{-}\left.\dfrac{\text{d}\theta}{\text{d}\eta}\right|_{\eta_{b}^{-}}\left(\mathcal{P}_{n}^{-}-\mathcal{P}_{n}^{-}\right)\phi_{b}+\\
\bar{D}_{R}\left(Z_{2}\right)\dfrac{\mathcal{P}_{d}^{-}}{\mathcal{P}_{d}^{+}}\left.\dfrac{\text{d}\phi}{\text{d}\eta}\right|_{\eta_{b}^{-}},\label{eq:mag field dev jump}
\end{multline}
where $\theta_{b}$, $\phi_{b}$ refer to their value at the ablated
border, and the superscripts ``$^{+}$'' and ``$^{-}$'' on the
transport polynomials refer to their evaluation to the left $\left(x_{e}^{-};Z_{2}\right)$
and to the right $\left(x_{e}^{+};Z_{1}\right)$ side of the ablated
border, respectively. Since $\text{Le}_{\text{m}}$ is typically large
in MagLIF implosions, the first term in the right-hand side of Eq.
\eqref{eq:mag field dev jump} is usually predominant. It is therefore
the variation of the Nernst convection velocity with $Z$, that is,
the factor $\left(\mathcal{P}_{n}^{-}-\mathcal{P}_{n}^{+}\right)$,
what originates a steep thin layer at the ablated border where the
magnetic field value changes drastically. 

The system of governing equations consists thereby of Eqs. \eqref{eq:ion continuity in T self-similar},
\eqref{eq:induction self-similar}, together with boundary conditions
\eqref{eq:BC for T}, \eqref{eq:BC for B} and jump conditions \eqref{eq:temperature dev jump},
\eqref{eq:mag field dev jump}. It depends on three free parameters:
the atomic number of the liner material $Z_{2}$, the magnetic Lewis
number $\text{Le}_{\text{m}}$ and the electron Hall parameter of
the unperturbed plasma $x_{e0}$. Typically, high magnetic Lewis numbers,
$\text{Le}_{\text{m}}\sim100-10^{6}$, are attained throughout a MagLIF
implosion\cite{velikovich2015magnetic}. Notice that the system of
governing equations derived in Ref. \onlinecite{garcia2017mass} is
recovered if we take $Z_{2}=Z_{1}=1$. Our final aim is to study mass
ablation, thermal energy and magnetic flux losses as a function of
these parameters.

\subsection{Mass ablation, thermal energy and magnetic flux losses}

An asymptotic analysis of the Eqs. \eqref{eq:ion continuity in T self-similar}
and \eqref{eq:induction self-similar} performed for $\eta\ll1$ reveals
that the temperature and magnetic field profiles close to the liner
take the form
\begin{equation}
\left.\theta\right|_{\eta\ll1}=s_{\theta}\eta^{2/5},\quad\left.\phi\right|_{\eta\ll1}=\text{Le}_{\text{m}}s_{\phi}\eta^{8/5}.\label{eq:profiles close to the liner}
\end{equation}
The parameters $s_{\theta}$ and $s_{\phi}$ depend on $\text{Le}_{\text{m}}$,
$x_{e0}$ and $Z_{2}$, and are obtained by solving the complete problem
with the boundary conditions far from the liner. 

We define mass ablation $m$ and magnetic flux losses $\Phi$ per
unit area in the hot spot as

\begin{equation}
m=\int_{0}^{x_{b}}\rho\text{d}x,\label{eq:mass increase definition}
\end{equation}
\begin{equation}
\Phi=\int_{0}^{\infty}\left(B_{0}-B\right)\text{d}x,\label{eq:magnetic flux definition}
\end{equation}
 Other quantities of interest are the thermal energy losses $\mathcal{E}$
and magnetic flux losses $\tilde{\Phi}$ in the fuel, defined as 
\begin{equation}
\mathcal{E}=\int_{0}^{\infty}\dfrac{p_{0}}{\gamma-1}\text{d}x-\int_{x_{b}}^{\infty}\dfrac{p_{0}}{\gamma-1}\text{d}x=\dfrac{p_{0}}{\gamma-1}x_{b},\label{eq:energy initial hot plasma definition}
\end{equation}
\begin{equation}
\tilde{\Phi}=\int_{0}^{\infty}B_{0}\text{d}x-\int_{x_{b}}^{\infty}B\text{d}x.\label{eq:magnetic flux initial hot plasma definition}
\end{equation}

Following a similar analysis as performed in Ref. \onlinecite{garcia2017mass},
these quantities are related to $s_{\theta}$, $s_{\phi}$ and $\eta_{b}$
as 
\begin{equation}
\dfrac{m}{\rho_{0}\sqrt{\kappa_{0}t}}=\dfrac{Z_{2}}{Z_{1}}\dfrac{Z_{1}+1}{Z_{2}+1}\bar{K}_{R}\left(Z_{2}\right)\dfrac{4}{5}s_{\theta}^{5/2},\label{eq:mass normalized}
\end{equation}
\begin{equation}
\dfrac{\Phi}{B_{0}\sqrt{\kappa_{0}t}}=\bar{D}_{R}\left(Z_{2}\right)\dfrac{16}{5}\dfrac{s_{\phi}}{s_{\theta}^{3/2}},\label{eq:magnetic flux normalized}
\end{equation}
\begin{equation}
\dfrac{\left(\gamma-1\right)\mathcal{E}}{p_{0}\sqrt{\kappa_{0}t}}=\eta_{b},\label{eq:energy initial hot palsma}
\end{equation}
\begin{equation}
\dfrac{\tilde{\Phi}}{B_{0}\sqrt{\kappa_{0}t}}=2s_{b},\label{eq:magnetic flux lossesinitial hot plasma}
\end{equation}
with 
\begin{equation}
s_{b}=\left.\left(\bar{K}_{R}\mathcal{P}_{n}\mathcal{P}_{c}\theta^{5/2}\dfrac{\text{d}\theta}{\text{d}\eta}\phi+\dfrac{\mathcal{P}_{d}\bar{D}_{R}}{\text{Le}_{\text{m}}\theta^{3/2}}\dfrac{\text{d}\phi}{\text{d}\eta}\right)\right|_{\eta=\eta_{b}^{-}}.\label{eq:parameter sb}
\end{equation}
The magnetic flux in the fuel is thereby lost due to Nernst convection
and magnetic diffusion through the ablated border, the former being
predominant if $\text{Le}_{\text{m}}$ is large. The fuel thermal
energy is lost due to thermal conduction at the ablated border, proportional
to $\eta_{b}$, and is inverted into heating the ablated liner material.

\section{analysis and results without mass diffusion\label{sec:analysis-and-results}}

\begin{figure}
\includegraphics[scale=0.25]{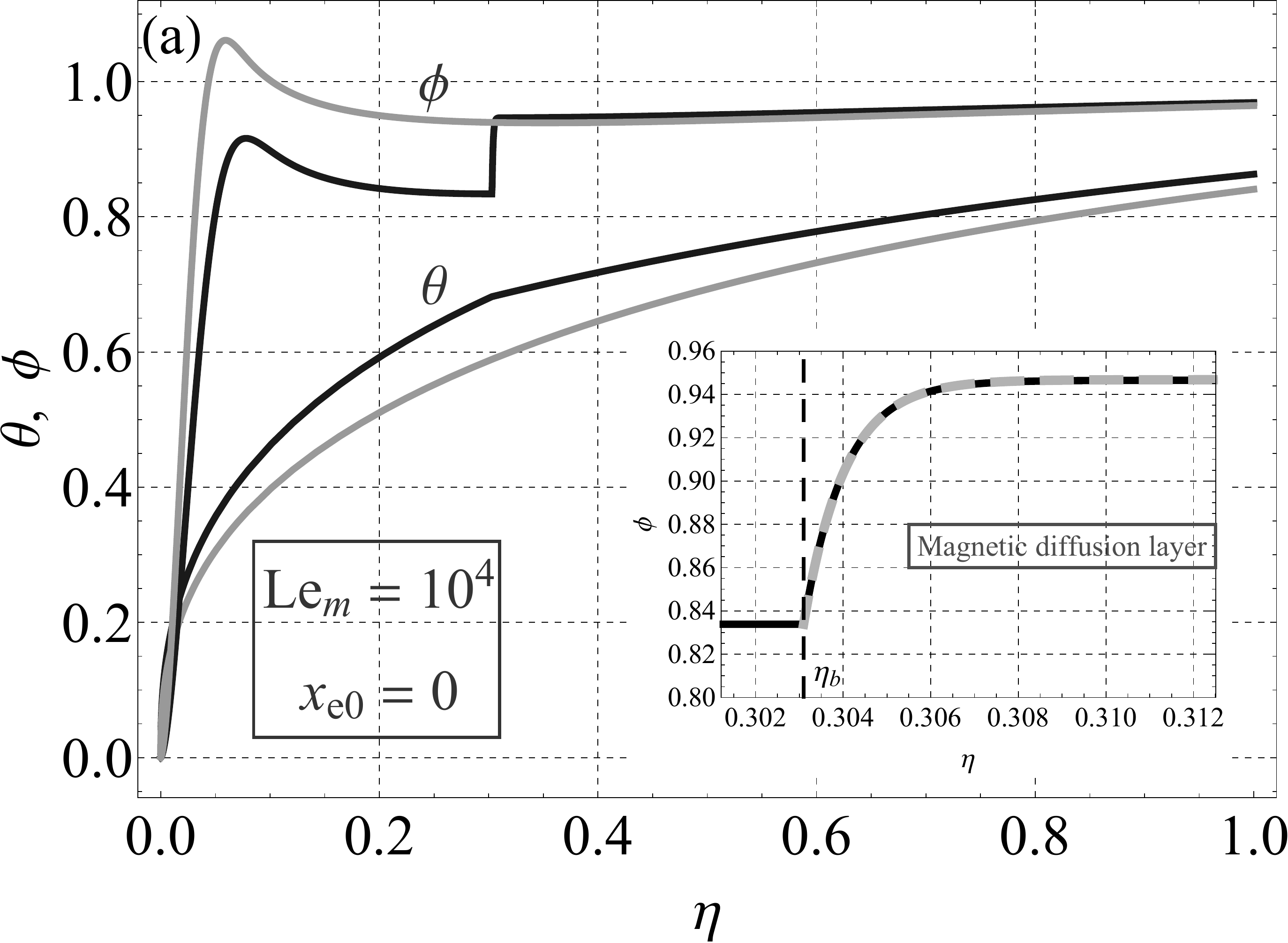}

\includegraphics[scale=0.25]{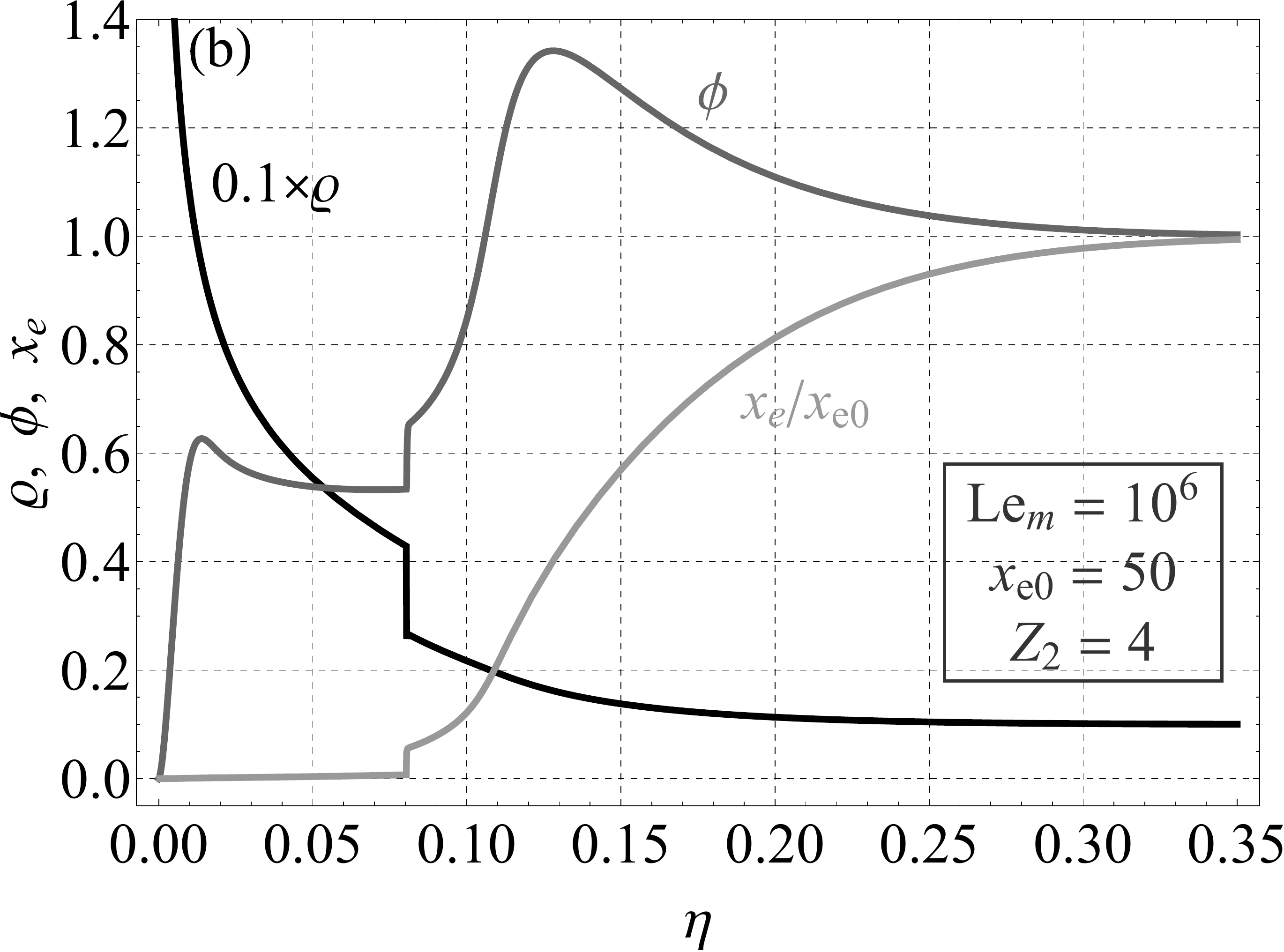}

\caption{\label{fig:No Diff Profiles}(a) Temperature $\theta$ and magnetic
field $\phi$ profiles for $\text{Le}_{\text{m}}=10^{4}$, $x_{e0}=0$
and $Z_{2}=1$ (gray) and $Z_{2}=4$ (black). The magnetic field diffusion
layer at $\eta=\eta_{b}$ is compared to the analytic expression Eq.
\eqref{eq:mgnetic field in the layer}, which is shown in dashed lines
in gray. (b) Mass density $\varrho$, magnetic field $\phi$ and Electron
Hall parameter $x_{e}$ profiles for $\text{Le}_{\text{m}}=10^{6}$,
$x_{e0}=50$ and $Z_{2}=4$. }
\end{figure}

Numerical computations of the governing equations are plotted in Fig.
\ref{fig:No Diff Profiles}. Temperature and magnetic field profiles
for the unmagnetized case and $\text{Le}_{\text{m}}=10^{4}$ are compared
in Fig. \ref{fig:No Diff Profiles}(a) for two different liner materials,
$Z_{2}=1$ (deuterium) and $Z_{2}=4$ (beryllium). It can be seen
that the plasma temperature at the hot spot is significantly higher
in the second case. The magnetic field, convected by the Nernst term
into the ablated material, is pushed against the liner and diffuses
in a thin layer adjacent to it. When the liner is made of beryllium,
the magnetic field experiments an abrupt decrease at the ablated border,
$\eta_{b}=0.303$; and as a consequence, the peak close to the liner
attains lower values. According to the jump condition \eqref{eq:mag field dev jump},
this decrease corresponds to the dependence of the Nernst convection
velocity on $Z$ through the polynomial $\mathcal{P}_{n}\left(x_{e};Z\right)$.
As seen in Fig. \ref{fig:Transport polynomials}(b), $\mathcal{P}_{n}\left(x_{e};4\right)>\mathcal{P}_{n}\left(x_{e};1\right)$
for any $x_{e}$, hence the Nernst convection velocity is stronger
in the ablated material. In order to conserve the amount of magnetic
field convected through the ablated border, $\phi$ is diffused in
a thin magnetic diffusion layer at the fuel side and its value is
reduced.

In Fig. \ref{fig:No Diff Profiles}(b), mass density, magnetic field
and magnetization profiles are shown for $\text{Le}_{\text{m}}=10^{6}$,
$x_{e0}=50$ and $Z_{2}=4$. In the magnetized region, the Nernst
term is suppressed and the magnetic filed moves frozen into the plasma.
Since the ablated material pushes the fuel inwards and compresses
it, the magnetic field presents a second peak at the right side of
the ablated border. At the left side, the electron Hall parameter
is relatively small and Nernst convection dominates. Again, a magnetic
diffusion layer takes place at the ablated border, $\eta_{b}=0.08$.
In this position, both the magnetization and mass density present
a discontinuity, given by Eqs. \eqref{eq:electron Hall jump} and
\eqref{eq:mass density jump}. The ablated border moves with negative
acceleration $g=\text{d}^{2}x_{b}/\text{d}t^{2}=-\eta_{b}\sqrt{\kappa_{0}}/4t^{3/2}<0$,
which implies that the light side of the ablated border (fuel) is
pushing backwards the heavy side (liner material). This magnetohydrodynamic
structure is susceptible to be Rayleigh-Taylor unstable, and would
deserve a more detailed study in a forthcoming paper. 

\subsection{Magnetic diffusion layer at the ablated border\label{subsec:Magnetic-diffusion-layer}}

In order to obtain the structure of the diffusion layer at the fuel
side of the ablated border, we expand the self-similar variable as
$\eta=\eta_{b}+\epsilon r$, with $\epsilon\ll1$, $r>0$. We assume
$\text{d}\phi/\text{d}r\sim O\left(1\right)$ in this layer and $\theta=\theta_{b}+O\left(\epsilon\right),V=\eta_{b}+O\left(\epsilon^{2}\right)$.
Inserting these ansätze into the induction equation \eqref{eq:induction self-similar}
yields to the leading order
\begin{equation}
\dfrac{\text{d}}{\text{d}r}\left(\mathcal{P}_{n}\phi\right)=-\dfrac{2}{\epsilon\text{Le}_{\text{m}}\eta_{b}\theta_{b}^{3/2}}\dfrac{\text{d}}{\text{d}r}\left(\mathcal{P}_{d}\dfrac{\text{d}\phi}{\text{d}r}\right),\label{eq:induction expanded}
\end{equation}
where the electron Hall parameter in the transport polynomials shall
be written as $x_{e}=x_{e0}\theta_{b}^{5/2}\phi$. We obtain then
the characteristic width of this layer 
\begin{equation}
\epsilon=\dfrac{2}{\mathcal{P}_{n}\left(0;Z_{1}\right)\text{Le}_{\text{m}}\eta_{b}\theta_{b}^{3/2}},\label{eq:width diffusion layer}
\end{equation}
which scales with the magnetic Lewis number as $\epsilon\sim O\left(1/\text{Le}_{\text{m}}\right)$.
This layer is thinner than the diffusion layer adjacent to the liner,
which scales as $O\left(1/\sqrt{\text{Le}_{\text{m}}}\right)$\cite{garcia2017mass}.

Equation \eqref{eq:induction expanded} stands for a nonlinear equation
of second order for $\phi$ that must be complemented with the boundary
conditions: $\phi\left(r=0\right)=\phi_{b}$ and the magnetic field
derivative given by Eq. \eqref{eq:mag field dev jump}. An analytical
solution can be obtained in the unmagnetized limit, $x_{e0}\ll1$,
in which case Eq. \eqref{eq:induction expanded} takes the form $\text{d}^{2}\phi/\text{d}r^{2}+\text{d}\phi/\text{d}r=0$,
and the solution reads
\begin{equation}
\phi\left(r\right)=\phi_{b}\left[\dfrac{\mathcal{P}_{n}^{-}-\mathcal{P}_{n}^{+}}{\mathcal{P}_{n}^{+}}\left(1-e^{-r}\right)+1\right].\label{eq:mgnetic field in the layer}
\end{equation}

This analytical expression is compared to the numerical solution in
Fig. \ref{fig:No Diff Profiles}(a), showing good agreement. Evaluating
this profile far from the layer, $r\rightarrow\infty$, allows to
obtain the magnetic field jump across it ($\phi^{r}$ to the right
and $\phi^{l}=\phi_{b}$ to the left), yielding: 
\begin{equation}
\dfrac{\phi^{r}}{\phi_{b}}=\dfrac{\mathcal{P}_{n}^{-}}{\mathcal{P}_{n}^{+}}>0.\label{eq:Bfield jump}
\end{equation}

\subsection{Analytic solution for large $Z_{2}$\label{subsec:Analytic-solution-for}}

In order to shed more light on the effect of the liner material, we
solve the governing equations \eqref{eq:ion continuity in T self-similar},
\eqref{eq:induction self-similar} in the limit of large liner atomic
number $Z_{2}$ and $x_{e0}\ll1$. When $Z_{2}$ is large, the conductivity
constant $\bar{K}$ in the ablated material is reduced, while the
diffusivity constant $\bar{D}$ is increased. In this limit, we can
write
\begin{equation}
\bar{K}_{R}\left(Z_{2}\gg1\right)=\dfrac{3.94}{Z_{2}}\equiv\epsilon_{K},\label{eq:KR Zgrande}
\end{equation}
\begin{equation}
\bar{D}_{R}\left(Z_{2}\gg1\right)=0.57Z_{2}\equiv\dfrac{\alpha}{\epsilon_{K}},\label{eq:DR Zgrande}
\end{equation}
with $\epsilon_{K}\ll1$ and $\alpha\equiv\alpha_{0}\left(\infty\right)\gamma_{0}\left(\infty\right)/\alpha_{0}\left(1\right)\gamma_{0}\left(1\right)\approx2.27$.
When $x_{e0}\ll1$, the continuity equation is uncoupled from the
induction equation, and can be solved independently. The transport
polynomials take the value $\mathcal{P}_{c}\left(0;Z\right)\approx1$,
$\mathcal{P}_{d}\left(0;Z\right)=1$, and $\mathcal{P}_{n}\left(0;1\right)\approx1.24$,
$\mathcal{P}_{n}\left(0;\infty\right)\approx1.83$. 

\subsubsection{Temperature profile}

In the ablated liner material, the small parameter $\epsilon_{K}$
can be absorbed in the continuity equation by scaling the independent
variable as $\eta=\sqrt{\epsilon_{K}}s$. Letting $\theta_{l}$ be
the temperature profile in this region, the continuity equation reads
\begin{equation}
s\dfrac{\text{d}\theta_{l}}{\text{d}s}+2\theta_{l}^{2}\dfrac{\text{d}}{\text{d}s}\left(\theta_{l}^{3/2}\dfrac{\text{d}\theta_{l}}{\text{d}s}\right)=0.\label{eq:ion continuity ablated material Zl large}
\end{equation}
The temperature profile close to the liner takes the form $\theta_{l}\left(s\ll1\right)=s_{\theta}^{\prime}s^{2/5}$,
with $s_{\theta}^{\prime}=s_{\theta}\epsilon_{K}^{1/5}$. The solution
has to satisfy the jump conditions at the ablated border. Scaling
its position as $s_{b}\equiv\eta_{b}/\sqrt{\epsilon_{K}}$, it is
determined by Eq. \eqref{eq:eta b}, which transforms into $\left.2\theta_{l}^{5/2}\text{d}\theta_{l}/\text{d}s\right|_{s_{b}}=s_{b}$.
At this position, the temperature must be continuous, $\theta_{l}\left(s_{b}\right)=\theta_{f}\left(\eta_{b}\right)$,
being $\theta_{f}\left(\eta\right)$ the temperature profile in the
fuel region. The temperature derivatives are linked through the jump
condition \eqref{eq:temperature dev jump}, giving $\left.\text{d}\theta_{f}/\text{d}\eta\right|_{\eta_{b}}=\sqrt{\epsilon_{K}}\theta_{lb}^{'}$,
where $\theta_{lb}^{\prime}\equiv\left.\text{d}\theta_{l}/\text{d}s\right|_{s_{b}}$.
If we assume that $\text{d}\theta_{l}/\text{d}s\sim O\left(1\right)$,
then the temperature derivative in the fuel is small of order $\sqrt{\epsilon_{K}}$,
and the solution will not differ much from $\theta_{f}\approx1$.
Therefore, to the leading order, we can impose that $\theta_{l}\left(s_{b}\right)=1$.
Consequently, the temperature profile at the ablated material $\theta_{l}$
can be solved independently of the fuel region in a small $\epsilon_{K}$
limit. The solution yields $s_{\theta}^{\prime}=1.21,$ $s_{b}=0.81=2\theta_{lb}^{\prime}$. 

In the fuel region, the continuity equation reads 
\begin{equation}
\eta\dfrac{\text{d}\theta_{f}}{\text{d}\eta}+2\theta_{f}^{2}\dfrac{\text{d}}{\text{d}\eta}\left(\theta_{f}^{3/2}\dfrac{\text{d}\theta_{f}}{\text{d}\eta}\right)=0.\label{eq:ion continuity ablated material Zl large-1}
\end{equation}
Since the temperature derivatives are small, we expand the solution
as $\theta_{f}=1-\sqrt{\epsilon_{K}}\theta_{f1}+o\left(\sqrt{\epsilon_{K}}\right).$
Inserting this ansatz in the previous equation and retaining the leading
order terms allows to obtain the first correction, giving $\theta_{f1}=\mathcal{C}\left[1-\text{Erf}\left(\eta/2\right)\right]$.
The constant of integration $\mathcal{C}$ is obtained by applying
the jump condition \eqref{eq:temperature dev jump}, yielding $\mathcal{C}=\sqrt{\pi}\theta_{lb}^{\prime}\approx0.72$.
The ion velocity and the Nernst velocity, proportional to the temperature
derivative, are also small of order $O\left(\sqrt{\epsilon_{K}}\right)$.

From these results, we infer that the ablated mass and energy losses
are written as

\begin{equation}
\dfrac{m}{\rho_{0}\sqrt{\kappa_{0}t}}=\dfrac{Z_{1}+1}{Z_{1}}\dfrac{4}{5}s_{\theta}^{\prime5/2}\sqrt{\epsilon_{K}}\approx\dfrac{5.09}{\sqrt{Z_{2}}},\label{eq:mass Zl large}
\end{equation}
\begin{equation}
\dfrac{\left(\gamma-1\right)\mathcal{E}}{p_{0}\sqrt{\kappa_{0}t}}=s_{b}\sqrt{\epsilon_{K}}\approx\dfrac{1.61}{\sqrt{Z_{2}}}.\label{eq:energy loss Zl large}
\end{equation}

To conclude, in a large $Z_{2}$ limit, the ablated border steps back,
the temperature profile is almost constant in the fuel and decreases
in the thin ablated liner layer whose width scales as $O\left(1/\sqrt{Z_{2}}\right)$,
and both mass ablation and energy losses are reduced when $Z_{2}$
increases. 

\subsubsection{Magnetic field profile}

We assume, as typically occurs in MagLIF, large $\text{Le}_{\text{m}}$.
Consequently, diffusion can be initially neglected in the fuel. The
magnetic field is convected by the plasma motion and the Nernst term,
being the latter predominant in the unmagnetized case. We introduce
$\delta_{l}=\mathcal{P}_{n}\left(0;\infty\right)-1\approx0.83$ and
$\delta_{f}=\mathcal{P}_{n}\left(0;1\right)-1\approx0.24$. The equation
governing the magnetic field in the fuel, $\phi_{f}$, reads then

\begin{equation}
\eta\dfrac{\text{d}\phi_{f}}{\text{d}\eta}+2\delta_{f}\sqrt{\epsilon_{K}}\theta_{lb}^{\prime}\dfrac{\text{d}}{\text{d}\eta}\left(\text{e}^{-\eta^{2}/4}\phi_{f}\right)=0.\label{eq:induction fuel Zl large}
\end{equation}
The solution satisfying $\phi_{f}\left(\infty\right)=1$, retaining
terms up to $O\left(\sqrt{\epsilon_{K}}\right)$, gives $\phi_{f}=1-\sqrt{\epsilon_{K}\pi}\delta_{f}\theta_{lb}^{\prime}\left[1-\text{Erf}\left(\eta/2\right)\right]$.
The magnetic field is barely perturbed since the convection velocity
is small, of order $O\left(\sqrt{\epsilon_{K}}\right)$, and only
the leading order $\phi_{f}\approx1$ will be retained. Close to the
ablated border, the thin diffusion layer described in Subsec. \ref{subsec:Magnetic-diffusion-layer}
takes place, its width given by Eq. \eqref{eq:width diffusion layer}:
$\epsilon\sim O\left(1/\text{Le}_{\text{m}}\sqrt{\epsilon_{K}}\right)$,
and the magnetic field in this layer drops from $\phi_{f}\approx1$
to $\phi_{b}=\left(1+\delta_{f}\right)/\left(1+\delta_{l}\right)=0.68$. 

In the ablated plasma liner, the induction equation shall be rewritten
as

\begin{equation}
-s\dfrac{\text{d}\phi_{l}}{\text{d}s}-2\delta_{l}\dfrac{\text{d}}{\text{d}s}\left(\theta_{l}^{5/2}\dfrac{\text{d}\theta_{l}}{\text{d}s}\phi_{l}\right)=\dfrac{2\alpha}{\text{Le}_{\text{m}}\epsilon_{K}^{2}}\dfrac{\text{d}}{\text{d}s}\left(\theta^{-3/2}\dfrac{\text{d}\phi_{l}}{\text{d}s}\right),\label{eq:induction liner Zl large}
\end{equation}
where we assume $\text{d}\phi_{l}/\text{d}s\sim O\left(1\right)$.
It can be seen that the magnetic Lewis number appropriate for this
region is $\text{Le}_{l}=\text{Le}_{\text{m}}\epsilon_{K}^{2}/\alpha$.
As discussed at the end of this section, we will assume that $\text{Le}_{l}$
is large too. Consequently, diffusion can be neglected in the main
part of the ablated material region, while it is confined to a thin
sub-layer close to the liner. Letting $\phi_{lo}$ denote the leading
order solution (outer solution) of Eq. \eqref{eq:induction liner Zl large},
and $w=2\theta_{l}^{5/2}\text{d}\theta_{l}/\text{d}s$, the former
satisfies
\begin{equation}
\dfrac{\text{d}\phi_{lo}}{\text{d}s}=-\delta_{l}\phi_{lo}\dfrac{\text{d}w/\text{d}s}{s+\delta_{l}w},\label{eq:outer solution}
\end{equation}
together with $\phi_{lo}\left(s_{b}\right)=\phi_{b}$. Solving this
equation allows to obtain the shape of this profile close to the liner
$\phi_{lo}\left(s\ll1\right)=s_{\phi_{o}}s^{-2/5}$, with $s_{\phi_{o}}=0.28$.
A thin sub-layer must then take place where the magnetic field diffuses
and drops to its value at the liner, $\phi_{l}\left(0\right)=0$.
In order to obtain the magnetic field in this sub-layer (inner solution
$\phi_{li}$), we expand Eq. \eqref{eq:induction liner Zl large}
in the variable $s=\epsilon_{B}z$, with $\epsilon_{B}\ll1$, and
assume $\text{d}\phi_{li}/\text{d}z\sim O\left(1\right)$. Requiring
diffusion to be important gives the width of this boundary layer
\begin{equation}
\epsilon_{B}=\sqrt{\dfrac{5}{2\delta_{l}s_{\theta}^{\prime}\text{Le}_{l}}},\label{eq:epsilonB}
\end{equation}
while the resulting equation for $\phi_{li}$ is 
\begin{equation}
\dfrac{\text{d}}{\text{d}z}\left(z^{2/5}\phi_{li}\right)+\dfrac{\text{d}}{\text{d}z}\left(\dfrac{1}{z^{3/5}}\dfrac{\text{d}\phi_{li}}{\text{d}z}\right)=0.\label{eq:eq for phili}
\end{equation}
The solution satisfying the boundary condition $\phi_{li}\left(0\right)=0$
is
\begin{equation}
\phi_{li}=Q\left[\int_{0}^{z}\hat{z}^{3/5}\exp\left(\hat{z}^{2}/2\right)\text{d}\hat{z}\right]\exp\left(-z^{2}/2\right),\label{eq:phi li}
\end{equation}
where the constant of integration $Q$ is obtained by matching the
inner solution $\phi_{li}\left(z\right)$ for $z\rightarrow\infty$
with the outer $\phi_{lo}\left(s\right)$ for $s\rightarrow0$, yielding
$Q=s_{\phi_{o}}\left(2\delta_{l}s_{\theta}^{\prime5}\text{Le}_{l}/5\right)^{1/5}$.
The solution in the ablated material region can be therefore expressed
by an inner-outer composite expansion\cite{van1975perturbation}:
$\phi_{l}\left(s\right)=\phi_{li}\left(s/\epsilon_{B}\right)+\phi_{lo}\left(s\right)-s_{\phi_{o}}s^{-2/5}$.
The parameter $s_{\phi}$ yields, in this double limit large $Z_{2}$
and large $\text{Le}_{\text{l}}$, $s_{\phi}=\epsilon_{K}^{6/5}s_{\phi_{o}}\delta_{l}s_{\theta}^{\prime5}/4\alpha\approx0.066\epsilon_{K}^{6/5}.$
From this analysis, we can derive the magnetic flux losses in the
hot spot and in the fuel, reading
\begin{equation}
\dfrac{\Phi}{B_{0}\sqrt{\kappa_{0}t}}=\dfrac{4}{5}\delta_{l}s_{\phi_{o}}s_{\theta}^{\prime7/2}\sqrt{\epsilon_{K}}\approx\dfrac{0.72}{\sqrt{Z_{2}}},\label{eq:mflhs large Zl}
\end{equation}
\begin{equation}
\dfrac{\tilde{\Phi}}{B_{0}\sqrt{\kappa_{0}t}}=2\left(1+\delta_{l}\right)\theta_{lb}^{\prime}\phi_{b}\sqrt{\epsilon_{K}}\approx\dfrac{2.01}{\sqrt{Z_{2}}}.\label{eq:mffs large Zl}
\end{equation}
In the large $\text{Le}_{l}$ limit, the main contribution to the
magnetic flux losses in the fuel corresponds to the magnetic field
convection through the ablated border due to the Nernst velocity,
that is, the first term in the right-hand side of Eq. \eqref{eq:parameter sb}.
The magnetic flux conservation both in the fuel and in the whole hot
spot (fuel plus ablated material) is improved when $Z_{2}$ increases,
following the same power law. This implies that the $64\%$ $\left[=100\times\left(1-\Phi/\tilde{\Phi}\right)\right]$
of the magnetic flux losses in the fuel are spent into magnetizing
the ablated material, while the remaining $36\%$ is lost through
dissipation at the liner. 

It is interesting to check the validity of this limit in realistic
MagLIF conditions. We have assumed that the ablated liner material
is fully ionized. The last ionization energy of aluminum, $Z_{2}=13$,
corresponds to 2.3 keV, which is attained by the end of the implosion.
At this stage, the magnetic Lewis number is large\cite{velikovich2015magnetic},
$\text{Le}_{\text{m}}\sim10^{6}$, therefore the assumption $\text{Le}_{l}\sim\text{Le}_{\text{m}}/Z_{2}^{2}\gg1$
is well satisfied.

\subsection{Results for mass ablation, thermal energy and magnetic flux losses}

\begin{figure}
\includegraphics[scale=0.25]{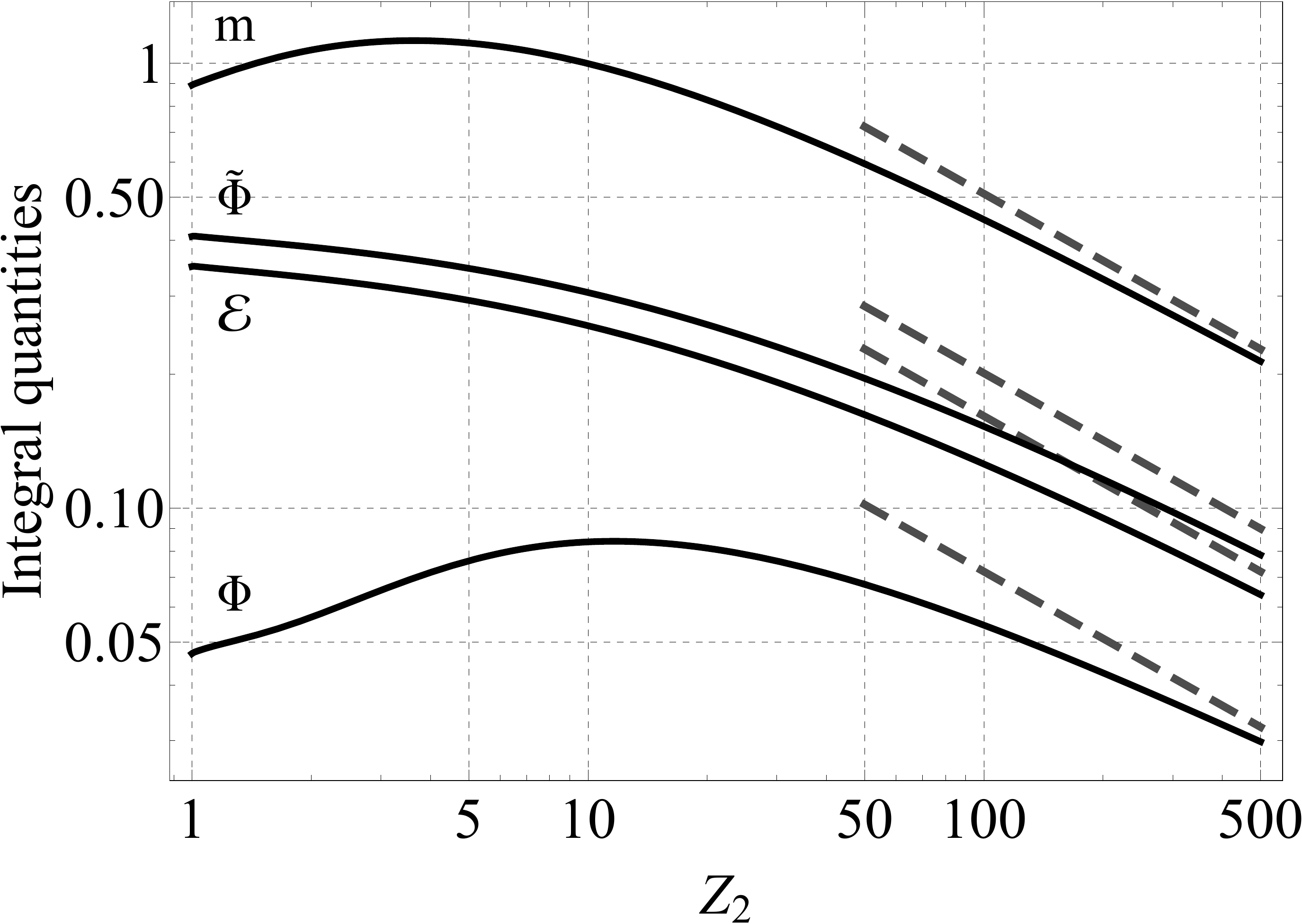}

\caption{\label{fig:Integral quantities vs Zl}Normalized mass ablation $m/\rho_{0}\sqrt{\kappa_{0}t}$,
thermal energy losses $\left(\gamma-1\right)\mathcal{E}/p_{0}\sqrt{\kappa_{0}t}=\eta_{b}$,
magnetic flux losses in the hot spot $\Phi/B_{0}\sqrt{k_{0}t}$ and
magnetic flux losses in the fuel $\tilde{\Phi}/B_{0}\sqrt{k_{0}t}$
for $\text{Le}_{\text{m}}=10^{9}$ and $x_{e0}=0$ as a function of
the atomic number of the liner $Z_{2}$. The dashed lines correspond
to the asymptotic laws derived in the large $Z_{2}$, $\text{Le}_{l}$
limit, equations \eqref{eq:mass Zl large}, \eqref{eq:energy loss Zl large},
\eqref{eq:mflhs large Zl} and \eqref{eq:mffs large Zl}.}
\end{figure}

In Fig. \ref{fig:Integral quantities vs Zl}, the integral quantities
mass ablation, energy losses and magnetic flux losses are shown as
a function of the atomic number of the liner $Z_{2}$ for $\text{Le}_{\text{m}}=10^{9}$
and $x_{e0}=0$. In order to obtain the integral quantities for any
$Z_{2}$, we have interpolated the transport coefficients given by
Braginskii\cite{braginskii1965transport} using splines of order 3,
and taking $Z^{-1}$ as the interpolation variable. The conservation
of both the thermal energy and the magnetic flux in the fuel is improved
when $Z_{2}$ increases. On the contrary, mass ablation and magnetic
flux losses in the hot spot attain a maximum for $Z_{2}=4$, $Z_{2}=12$,
respectively, where they take the value $1.12$ and $0.084$. Although
large $Z_{2}$ values are physically meaningless, they are plotted
to check the accuracy of the asymptotic laws derived in Subsec. \ref{subsec:Analytic-solution-for},
which are also shown in the figure. 

As stated in Subsec. \ref{subsec:Analytic-solution-for}, increasing
$Z_{2}$ lowers the thermal conductivity in the ablated material and
the ion and Nernst velocities in all the hot spot. Consequently, heat
flux at the ablated border is reduced, and less magnetic field is
convected through it by the Nernst term. Both $\mathcal{E}$ and $\tilde{\Phi}$
decrease then with $Z_{2}$. One would expect the same trend for $m$
and $\Phi$, as the convection velocity is reduced and less magnetic
field is accumulated at the liner. However, when $Z_{2}$ increases,
the ions coming from the liner are heavier, and the magnetic Lewis
number at the ablated liner, $\text{Le}_{l}\sim\text{Le}_{m}/Z_{2}^{2}$,
is reduced, that is, the plasma becomes less conductive. These effects
enhance $m$ and $\Phi$, respectively, and are predominant over the
reduction of convection velocity for small $Z_{2}$, while the latter
becomes more important for large $Z_{2}$, and therefore $m$ and
$\Phi$ present a maximum.

\begin{figure}
\includegraphics[scale=0.25]{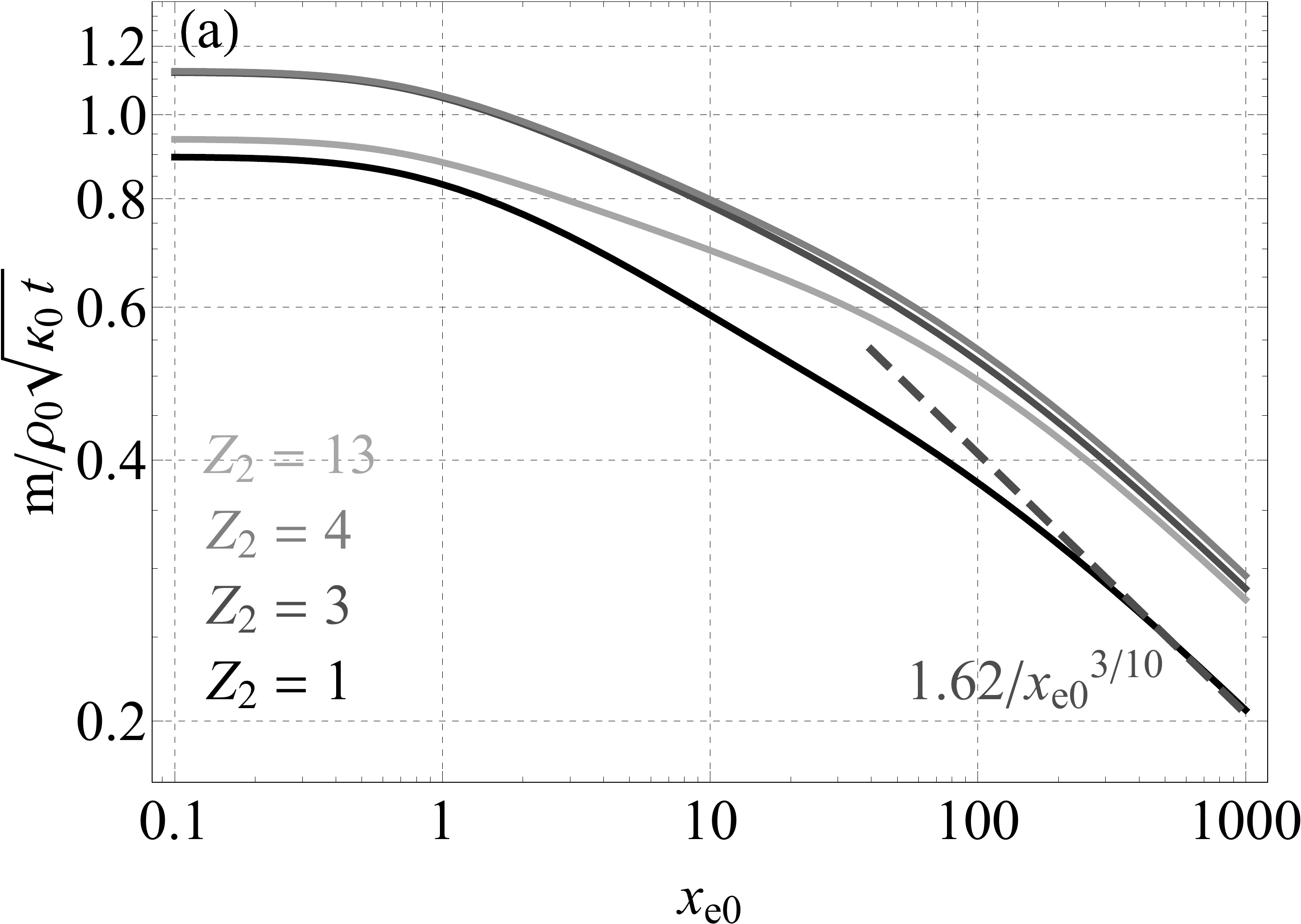}

\includegraphics[scale=0.25]{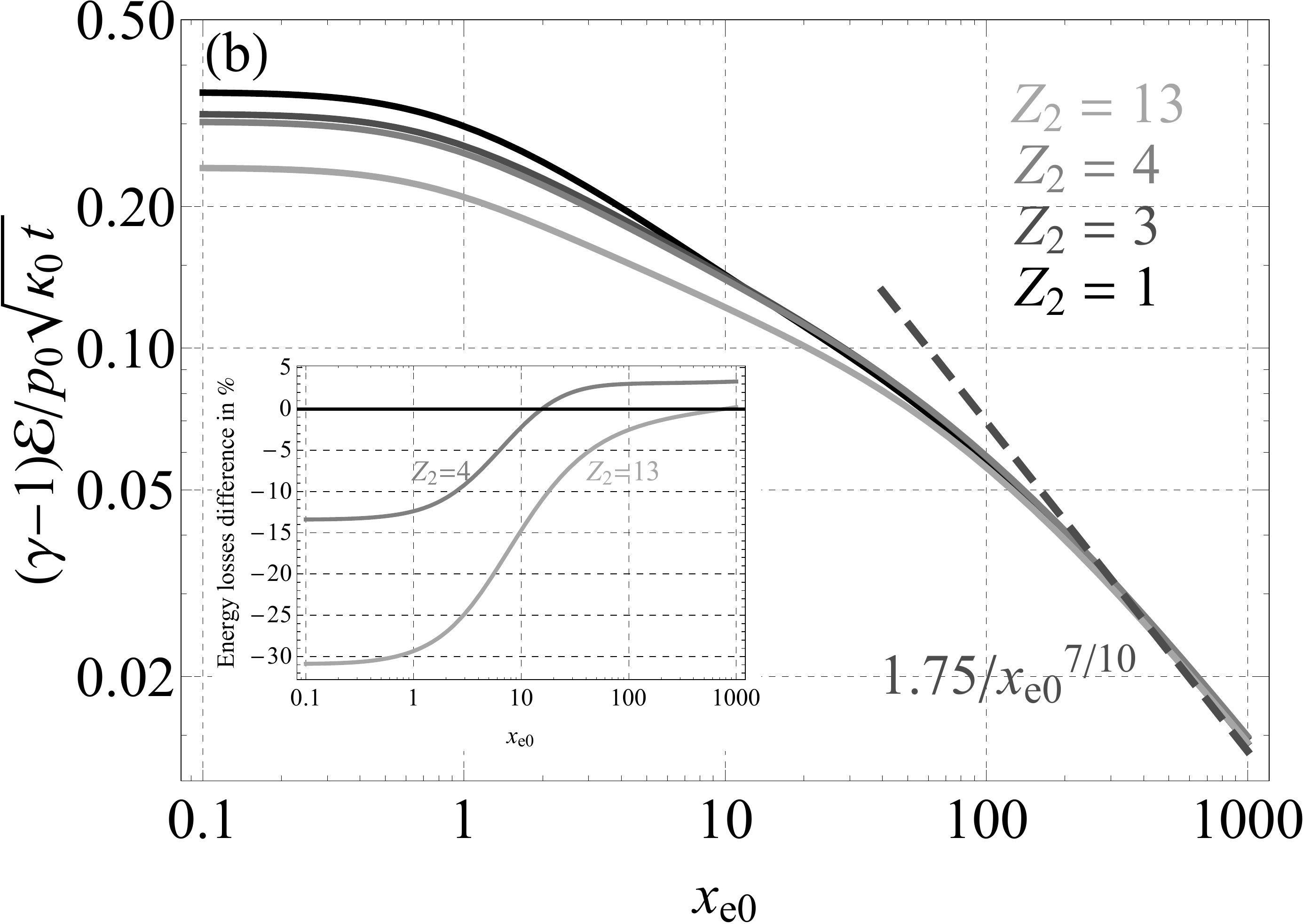}

\caption{\label{fig:Mass Energy}Normalized mass ablation $m/\rho_{0}\sqrt{\kappa_{0}t}$
(a) and thermal energy losses $\left(\gamma-1\right)\mathcal{E}/p_{0}\sqrt{\kappa_{0}t}=\eta_{b}$
(b) for different liner materials. The curves keep a constant $\beta=1000$
when the magnetization is increased. In (b), the difference in percentage
between the $Z_{2}=1$ and $Z_{2}=4$, $Z_{2}=13$ cases is plotted,
computed as $100\times\left(\left.\eta_{b}\right|_{Z_{2}=4}-\left.\eta_{b}\right|_{Z_{2}=1}\right)/\left.\eta_{b}\right|_{Z_{2}=1}$
and $100\times\left(\left.\eta_{b}\right|_{Z_{2}=13}-\left.\eta_{b}\right|_{Z_{2}=1}\right)/\left.\eta_{b}\right|_{Z_{2}=1}$,
respectively.}
\end{figure}

\begin{figure}
\includegraphics[scale=0.25]{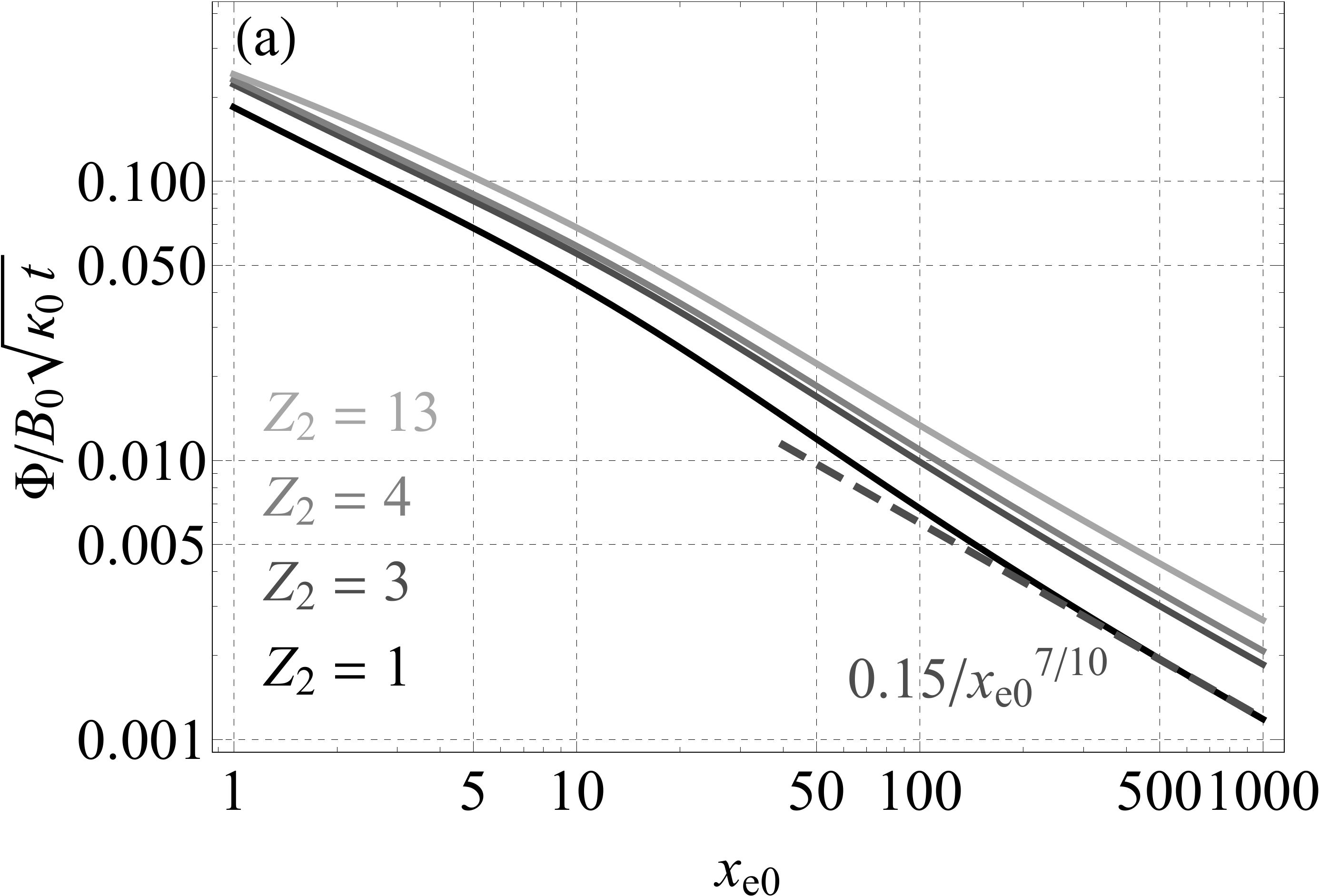}

\includegraphics[scale=0.25]{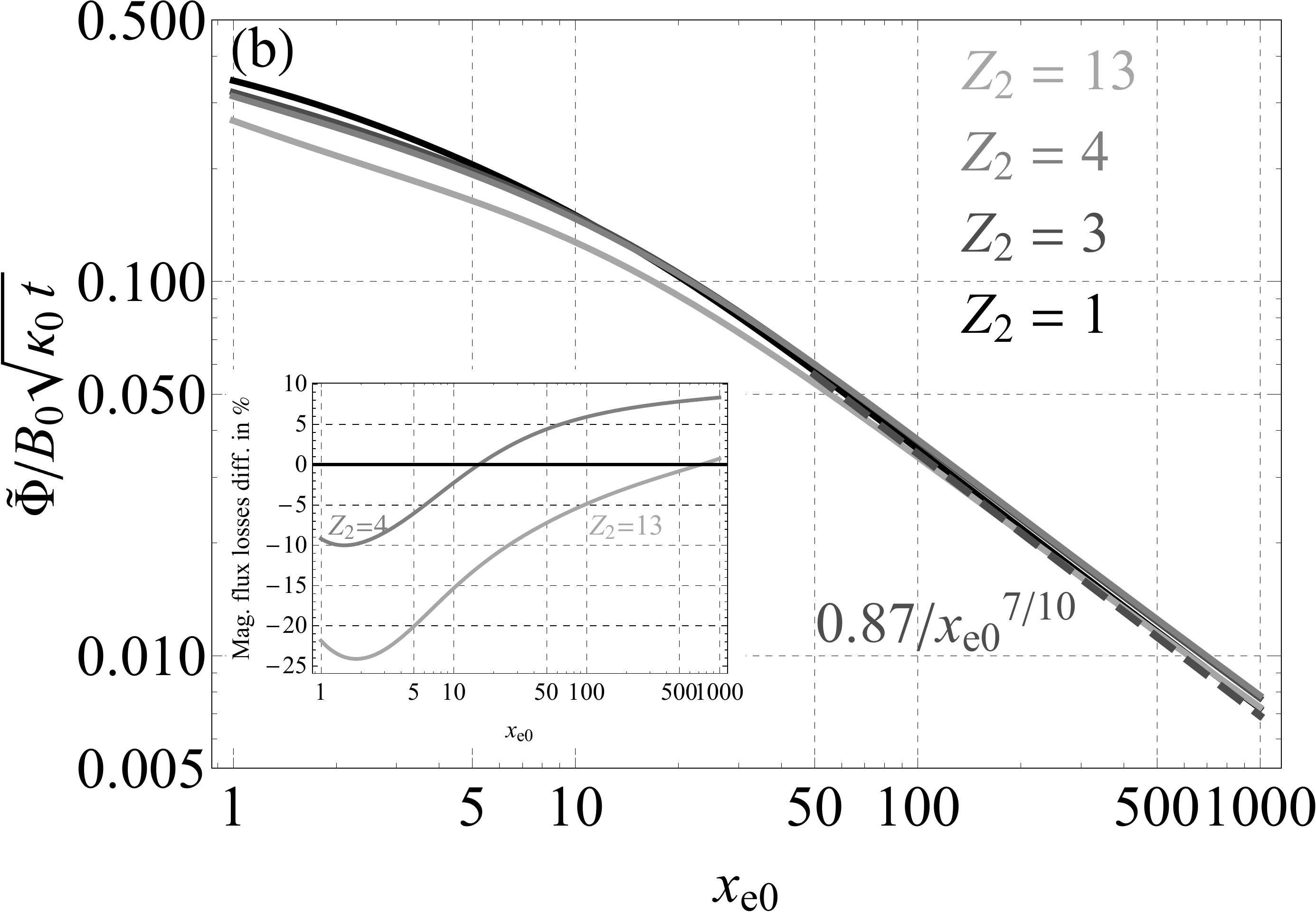}

\caption{\label{fig:Magnetic flux losses}Normalized magnetic flux losses computed
in: (a) the hot spot $\left(x>0\right)$, $\Phi/B_{0}\sqrt{\kappa_{0}t}$
and (b) the fuel $\left(x>x_{b}\right)$, $\tilde{\Phi}/B_{0}\sqrt{\kappa_{0}t}$,
for different liner materials. The curves keep a constant $\beta=1000$
when the magnetization is increased. In (b), the difference in percentage
between the $Z_{2}=1$ and $Z_{2}=4$, $Z_{2}=13$ cases is plotted,
computed as $100\times\left(\left.\tilde{\Phi}\right|_{Z_{2}=4}-\left.\tilde{\Phi}\right|_{Z_{2}=1}\right)/\left.\tilde{\Phi}\right|_{Z_{2}=1}$
and $100\times\left(\left.\tilde{\Phi}\right|_{Z_{2}=13}-\left.\tilde{\Phi}\right|_{Z_{2}=1}\right)/\left.\tilde{\Phi}\right|_{Z_{2}=1}$,
respectively. }
\end{figure}

The integral quantities mass ablation, thermal energy and magnetic
flux losses as a function of the fuel initial magnetization $x_{e0}$
and for different liner materials (lithium $Z_{2}=3$, beryllium $Z_{2}=4$
and aluminum $Z_{2}=13$) are shown in Figs. \ref{fig:Mass Energy}
and \ref{fig:Magnetic flux losses}. The curves are computed keeping
a large $\beta=1000$. In the same figures, asymptotic laws for large
$x_{e0}$ are plotted, some of which were obtained in Ref. \onlinecite{garcia2017mass}
and adjusted by numerical fitting. 

When the fuel is unmagnetized, low $x_{e0}$, the mass and energy
losses follow the same trend observed in Fig. \ref{fig:Integral quantities vs Zl}:
the ablated mass is enhanced for lithium and beryllium, it attains
a maximum, and then decreases for higher $Z_{2}$ (aluminum). The
energy losses, on their part, decrease monotonically with the atomic
number of the liner. Magnetizing the fuel reduces the mass ablated
and improves thermal insulation. The effect of $Z_{2}$ on thermal
insulation becomes less important when the fuel is magnetized. It
could even be inverted when the electron Hall parameter exceeds a
certain threshold, see Fig. \ref{fig:Mass Energy}(b), but it barely
enhances the energy losses by less than $5\%$. 

In Fig. \ref{fig:Magnetic flux losses}, the magnetic flux losses
in the hot spot and in the fuel are depicted. They are shown for $x_{e0}>1$,
since the magnetic Lewis number becomes relatively small when the
magnetization is further reduced keeping a constant $\beta=1000$,
and is out of the range of application to MagLIF. Nevertheless, the
arguments derived for the unmagnetized case still apply for moderate
$x_{e0}$. It can be seen that the magnetic flux losses in the hot
spot are enhanced when $Z_{2}$ is increased up to the aluminum value,
as the ablated material becomes more diffusive. This effect is preserved
for all values of $x_{e0}$. However, the magnetic flux conservation
in the fuel, Fig. \ref{fig:Magnetic flux losses}(b), follows a different
trend. It is improved with $Z_{2}$ for moderate magnetizations due
to the reduction of the Nernst velocity; but this effect diminishes
when the magnetization is increased. It can even be inverted for a
liner made of beryllium for $x_{e0}>15$, degrading the magnetic flux
conservation by less than $10\%$ compared to a liner made of dense
deuterium.

\section{governing equations with mass diffusion\label{sec:governing-equations-with-mass-transport}}

We formulate now the problem sketched in Fig. \ref{fig:Artistic scheme}
taking into account mass diffusion through the ablated border. The
hot spot is modeled as a two ion species plasma composed by deuterium
(fuel) and liner material, and we assume that every ion species is
fully ionized. The hydrodynamic description of a multiple species
plasma, including a closed expression for the transport terms, is
derived in Refs. \onlinecite{simakov2016hydrodynamicI,simakov2016hydrodynamicII}.
We follow the notation therein, therefore the fuel (light species)
is labeled with the subscript $1$, the liner material (heavy species)
with $2$ and the electrons with $e$. We name $n_{k}$, with $k=\left\{ 1,2,e\right\} $,
the particle number density, $\rho_{k}$ is the mass density, with
$\rho_{k}=m_{k}n_{k}$, and $m_{k}\approx2Z_{k}m_{p}$ when $k$ refers
to an ion species, being $Z_{k}$ its atomic number and $m_{p}$ the
proton mass. We assume same temperature for electrons and ions $T_{e}=T_{i}=T$.
The partial pressure is $p_{k}=n_{k}T$, and $\vec{v}_{k}$ refers
to the flow velocity. We define the plasma density $\rho=\sum_{k}\rho_{k}\approx\rho_{1}+\rho_{2}$,
the total ion number density $n_{i}=n_{1}+n_{2}$, the total ion pressure
$p_{i}=p_{1}+p_{2}=n_{i}T$ and the total plasma pressure $p=p_{i}+p_{e}$.
Note that in Refs. \onlinecite{simakov2016hydrodynamicI,simakov2016hydrodynamicII},
$p$ denotes the total ion pressure, yet, we choose to use $p$ for
the total plasma pressure to be consistent with the formulation in
the first part of this paper, which at the same time inherits the
original formulation in Ref. \onlinecite{garcia2017mass}. The plasma
ion velocity is defined as $\vec{v}=\left(\rho_{1}\vec{v}_{1}+\rho_{2}\vec{v}_{2}\right)/\rho$,
and consequently the electron and ion drift velocities are expressed
as $\vec{u}_{k}=\vec{v}_{k}-\vec{v}$. Again, although $\vec{u}$
refers to the plasma ion velocity and $\vec{v}_{k}$ to the drift
velocities in Refs. \onlinecite{simakov2016hydrodynamicI,simakov2016hydrodynamicII},
we choose to swap them and denote the plasma ion velocity as $\vec{v}$
to keep consistent with the first part of this paper. Finally, $x_{k}=n_{k}/n_{i}$
is the number density fraction of the ion species $k$, such that
$x_{1}+x_{2}=1$; $y=\rho_{1}/\rho$ is the mass concentration of
the fuel and $1-y$ is the mass concentration of the liner material. 

As commented in the introduction, we have to restrict the fuel liner
mixing analysis to the unmagnetized plasma case, where the plasma
motion is uncoupled from the magnetic field evolution. Studying mass
diffusion in a magnetized plasma would require first to extend the
transport theory of multi-component plasmas\cite{molvig2014classical,simakov2016hydrodynamicI,simakov2016hydrodynamicII}
to take into account the effect of magnetic fields, which is out of
the scope of this paper. In the absence of magnetic field, the Ampère's
law, $\vec{0}=4\pi\vec{j}+\partial\vec{E}/\partial t$, drives the
electric field $\vec{E}$ to the value set by the ambipolarity condition
$\vec{j}=0$, with $\vec{j}=e\left(Z_{1}n_{1}\vec{u}_{1}+Z_{2}n_{2}\vec{u}_{2}-n_{e}\vec{u}_{e}\right)$
being the plasma current. As a consequence of the planar geometry
and the absence of currents, the electron and ions velocities as well
as the rest of the velocities introduced only present a streamwise
component, and will be treated as scalars: $v_{k},v$ and $u_{k}$. 

The evolution of the plasma density $\rho$, pressure $p$, velocity
$v$ and fuel concentration $y$ in the hot spot, $x>0$, is given
by the total ion continuity, plasma momentum conservation, energy
conservation and fuel continuity equations, which in planar geometry
and low Mach and high $\beta$ limit yield 

\begin{equation}
\dfrac{\partial\rho}{\partial t}+\dfrac{\partial}{\partial x}\left(\rho v\right)=0,\label{eq:ion continuity dif}
\end{equation}

\begin{equation}
\dfrac{\partial p}{\partial x}=0,\label{eq:total momentum dif}
\end{equation}

\begin{multline}
\dfrac{3}{2}\dfrac{\partial p}{\partial t}+\dfrac{\partial}{\partial x}\left[\dfrac{5}{2}pv+q_{e}+q_{i}\right.\\
\left.+\dfrac{5}{2}\left(p_{1}u_{1}+p_{2}u_{2}+p_{e}u_{e}\right)\right]=E\cdot j\approx0,\label{eq:total energy dif}
\end{multline}
\begin{equation}
\rho\dfrac{\partial y}{\partial t}+\rho v\dfrac{\partial y}{\partial x}=-\dfrac{\partial}{\partial x}\left(\rho yu_{1}\right),\label{eq:Difusion dif}
\end{equation}
where $q_{e}$ and $q_{i}$ are the electron and ion heat fluxes,
respectively. The momentum conservation equation \eqref{eq:total momentum dif}
reduces to isobaricity, $p=p_{0}$. In addition to these equations,
we impose plasma quasi-neutrality, $n_{e}=Z_{1}n_{1}+Z_{2}n_{2}$.
Besides, from the definition of plasma ion velocity, we have $\rho_{1}u_{1}+\rho_{2}u_{2}=0$,
and the aforementioned ambipolarity condition gives $Z_{1}n_{1}u_{1}+Z_{2}n_{2}u_{2}-n_{e}u_{e}=0$.
Using $n_{1}=\rho_{1}/m_{1}=\rho y/m_{1}$, any ion species or electron
densities and drift velocities can be expressed in terms of $\rho$,
$y$ and $u_{1}$, uniquely. The equation of state can therefore be
written as 
\begin{equation}
p=p_{0}=\rho T\left[\left(1+Z_{1}\right)\dfrac{y}{m_{1}}+\left(1+Z_{2}\right)\dfrac{1-y}{m_{2}}\right].\label{eq:eos diff}
\end{equation}

With the isobaric assumption, plasma energy equation \eqref{eq:total energy dif}
can be integrated once, giving an explicit expression for the plasma
velocity
\begin{equation}
v=-\dfrac{2}{5p_{0}}\left[q_{e}+q_{i}+\dfrac{5}{2}\left(p_{1}u_{1}+p_{2}u_{2}+p_{e}u_{e}\right)\right].\label{eq:v diff}
\end{equation}
Expressing the plasma density $\rho$ in terms of $T$ and $y$ by
means of Eq. \eqref{eq:eos diff}, and making use of Eq. \eqref{eq:v diff},
we can reduce the governing equations to a system of two equations,
\eqref{eq:ion continuity dif} and \eqref{eq:Difusion dif}, for $T\left(x,t\right)$
and $y\left(x,t\right)$. To close this system, we require the expressions
for the electron and ion heat fluxes, $q_{e}$ and $q_{i}$, and the
fuel drift velocity $u_{1}$. These relations are given in Refs. \onlinecite{simakov2016hydrodynamicI,simakov2016hydrodynamicII},
and its derivation is briefly summarized in the Appendix \ref{sec:Review-of-transport}.
We make therefore use of Eqs. \eqref{eq:qe dif 2}, \eqref{eq:qi bis}
and \eqref{eq:u1} to relate $q_{e}$, $q_{i}$ and $u_{1}$ to $T$
and $y$. Notice that the order of the system of differential equations
is four. As boundary conditions, we impose that far from the liner,
$x\rightarrow\infty$, we recover the initial state, $T=T_{0}$ and
$y=1$, while at the liner - hot spot interface, $x=0$, we have $T=0$
and $y=0$.

\subsection{Normalization and self-similarity}

Similarly to Subsec. \ref{subsec:Normalization-and-self-similarit},
we normalize temperature and plasma density with their initial value
in the fuel: $\theta=T/T_{0}$ and $\varrho=\rho/\rho_{0}$. Again,
the solution is sought under the form of a self-similar diffusive
wave\cite{zel2002physics}. We introduce the independent self-similar
variable $\eta=x/\sqrt{\kappa_{0}t}$, with 
\begin{equation}
\kappa_{0}\equiv\dfrac{2\bar{K}_{e}\left(Z_{1}\right)T_{0}^{7/2}}{5p_{0}}\approx\dfrac{40,500}{\log\Lambda_{ee}}\dfrac{T_{0}^{5/2}}{\rho_{0}}\text{ cm}^{2}/\text{sec},\label{eq:kappa 0}
\end{equation}
being the thermal diffusivity as likewise defined in Eq. \eqref{eq:k0 no diff}.
Consistently, the plasma velocity and the fuel drift velocities are
scaled as
\begin{equation}
V\left(\eta\right)=2\sqrt{\dfrac{t}{\kappa_{0}}}v\left(x,t\right),\qquad U_{1}\left(\eta\right)=2\sqrt{\dfrac{t}{\kappa_{0}}}u_{1}\left(x,t\right).\label{eq:velocities nromalized dif}
\end{equation}

The governing equations \eqref{eq:ion continuity dif} and \eqref{eq:Difusion dif}
are rewritten as 
\begin{equation}
\left(V-\eta\right)\dfrac{\text{d}\varrho}{\text{d}\eta}+\varrho\dfrac{\text{d}V}{\text{d}\eta}=0,\label{eq:cont self-similar dif}
\end{equation}
\begin{equation}
\varrho\left(V-\eta\right)\dfrac{\text{d}y}{\text{d}\eta}=-\dfrac{\text{d}}{\text{d}\eta}\left(\varrho yU_{1}\right).\label{eq:eq dif self-similar}
\end{equation}

The dimensionless density $\varrho$ can be related to $\theta$ and
$y$ through the equation of state \eqref{eq:eos diff}, yielding
\begin{equation}
\varrho=\dfrac{1}{\theta\left(y+\dfrac{Z_{2}+1}{Z_{1}+1}\dfrac{1-y}{\mu}\right)},\label{eq:varrho}
\end{equation}
with $\mu=m_{2}/m_{1}$. The dimensionless plasma velocity is obtained
from Eq. \eqref{eq:v diff}, reading

\begin{multline}
V=2\left[\dfrac{\bar{K}_{e}\left(Z_{\text{eff}}\right)+\bar{K}_{i}\left(\upsilon\right)}{\bar{K}_{e}\left(Z_{1}\right)}\right]\theta^{5/2}\dfrac{\text{d}\theta}{\text{d}\eta}-\\
f_{\nu}\dfrac{y}{\left(Z_{1}+1\right)y+\left(Z_{2}+1\right)\dfrac{1-y}{\mu}}V_{1},\label{eq:v diff1}
\end{multline}
with $f_{\nu}=\left[2\hat{D}_{T1}^{\prime}/5\hat{\Delta}_{11}^{\prime}x_{1}+Z_{1}+1-\left(Z_{2}+1\right)/\mu\right]$.
Finally, the normalized fuel drift velocity is recovered from Eq.
\eqref{eq:u1}, and gives
\begin{equation}
U_{1}=-\dfrac{2}{\text{Le}}\dfrac{\theta^{7/2}}{y}\left[\left(\mathcal{D}_{c}+\mathcal{D}_{p}\right)\dfrac{\text{d}y}{\text{d}\eta}+\left(\mathcal{D}_{T_{e}}+\mathcal{D}_{T_{i}}\right)\dfrac{\text{d}\log\theta}{\text{d}\eta}\right],\label{eq:U1}
\end{equation}
where $\text{Le}$ stands for the Lewis number, typically defined
in mass transfer problems as the ratio between thermal and mass diffusivities,
$\kappa_{0}$ and $\nu_{0}$, respectively\cite{white1988heat}. The
latter corresponds to the characteristic value of the coefficient
relating fuel diffusion velocity and fuel mass concentration gradient
in Eq. \eqref{eq:u1}, and reads 

\begin{multline}
\nu_{0}\equiv\dfrac{2n_{1}}{\nu_{11}T^{3/2}}\dfrac{Z_{2}+1}{Z_{1}+1}\hat{\Delta}_{11}^{\prime}\left(0\right)\dfrac{T_{0}^{5/2}}{\rho_{0}}\approx\\
\dfrac{1,493}{\log\Lambda_{11}}P_{\mu}\dfrac{Z_{2}+1}{Z_{2}^{2}}\dfrac{T_{0}^{5/2}}{\rho_{0}}\text{ cm}^{2}/\text{sec},\label{eq:D0}
\end{multline}
with $P_{\mu}$ being an increasing function of the ion mass ratio
$\mu$, given in Eq. \eqref{eq:Pmu}. It ranges from $0.595$ for
$\mu=1$ to $1.20$ for $\mu\rightarrow\infty$. 

The Lewis number is then

\begin{equation}
\text{Le}\equiv\dfrac{\kappa_{0}}{\nu_{0}}=\dfrac{1}{5\sqrt{2}}\dfrac{\gamma_{0}\left(Z_{1}\right)Z_{1}^{3}}{Z_{2}+1}\dfrac{1}{\hat{\Delta}_{11}^{\prime}\left(0\right)}\sqrt{\dfrac{m_{1}}{m_{e}}}\approx\dfrac{27}{P_{\mu}}\dfrac{Z_{2}^{2}}{Z_{2}+1}.\label{eq:Lewis}
\end{equation}
Notice that it only depends on the liner atomic number, and its value
is shown in Table \ref{tab:Lewis} for typical liner materials. The
Lewis number is relatively large and becomes larger for higher $Z_{2}$,
which implies that mass transport is less effective than heat transport.
The motion of the plasma in the hot spot is therefore governed by
the heat wave, and mass diffusion is confined within a thin layer
placed at the ablated border. 

The terms $\mathcal{D}_{c}$, $\mathcal{D}_{p}$, $\mathcal{D}_{T_{e}}$
and $\mathcal{D}_{T_{i}}$ are the dimensionless coefficients for
concentration gradient diffusion, baro-diffusion, electron thermo-diffusion
and ion thermo-diffusion, respectively,

\begin{equation}
\mathcal{D}_{c}\equiv\dfrac{\hat{\Delta}_{11}^{\prime}\left(y\right)}{\hat{\Delta}_{11}^{\prime}\left(0\right)}\dfrac{1-y+\dfrac{Z_{1}+1}{Z_{2}+1}\mu y}{\left(1-y+\mu y\right)^{2}},\label{eq:Dc}
\end{equation}

\begin{multline}
\mathcal{D}_{p}\equiv\mathcal{D}_{c}y\left(1-y\right)\times\\
\dfrac{\mu\left(Z_{2}-Z_{1}\right)^{2}}{Z_{2}\left(Z_{2}+1\right)\left(1-y+\dfrac{Z_{1}}{Z_{2}}\mu y\right)\left(1-y+\dfrac{Z_{1}+1}{Z_{2}+1}\mu y\right)},\label{Dp}
\end{multline}

\begin{multline}
\mathcal{D}_{T_{e}}\equiv\dfrac{\hat{\Delta}_{11}^{\prime}\left(y\right)}{\hat{\Delta}_{11}^{\prime}\left(0\right)}y\left(1-y\right)\left(1-y+\dfrac{Z_{1}+1}{Z_{2}+1}\mu y\right)\times\\
\dfrac{Z_{1}}{Z_{2}}\dfrac{Z_{2}-Z_{1}}{\left(1-y+\mu y\right)\left(1-y+\dfrac{Z_{1}^{2}}{Z_{2}^{2}}\mu y\right)}\beta_{0}\left(x_{1}\right),\label{DTe-1}
\end{multline}
\begin{equation}
\mathcal{D}_{T_{i}}\equiv\dfrac{\hat{D}_{T1}^{\prime}\left(y\right)}{\hat{\Delta}_{11}^{\prime}\left(0\right)}\dfrac{1}{\mu}\left(1-y+\dfrac{Z_{1}+1}{Z_{2}+1}\mu y\right).\label{eq:DTi}
\end{equation}
Notice that, as a consequence of isobaricity, both the electron and
ion pressure gradient terms in Eq. \eqref{eq:d1 expression 3} can
be expressed in terms of fuel mass concentration gradients. The sum
of both effects is accounted for in $\mathcal{D}_{p}$, and is denoted
hereinafter as baro-diffusion.

\begin{figure}
\includegraphics[scale=0.25]{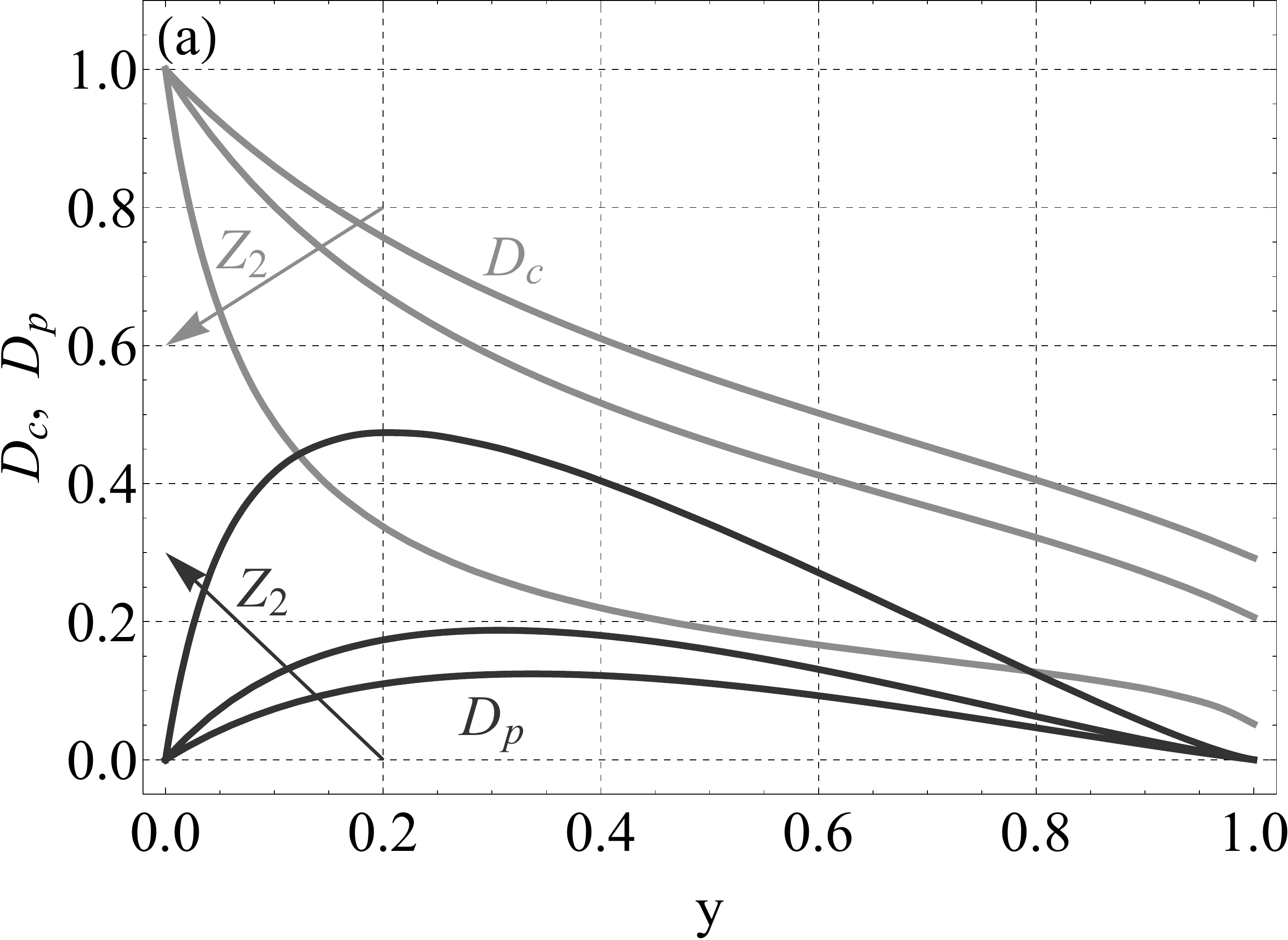}

\includegraphics[scale=0.25]{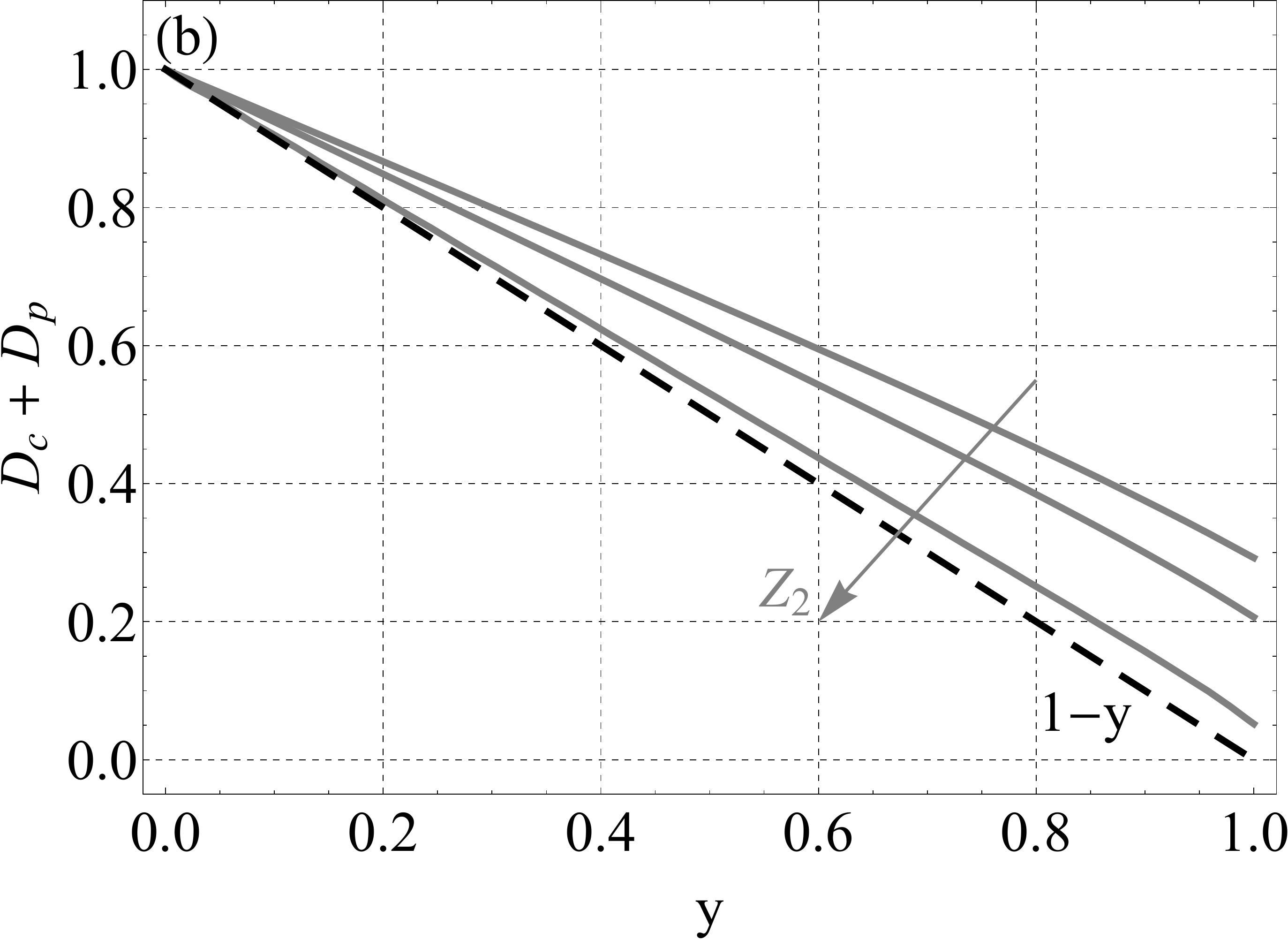}

\caption{\label{fig:Coeff Dc Dp}Concentration gradient diffusion coefficient,
$\mathcal{D}_{c}$, and baro-diffusion coefficient, $\mathcal{D}_{p}$,
for $Z_{2}=3$ (lithium), $Z_{2}=4$ (beryllium) and $Z_{2}=13$ (aluminum),
plotted separately in (a) and summed in (b)\@.}
\end{figure}

As will be explained in Subsec. \ref{subsec:Mass-diffusion-boundary},
concentration gradient diffusion and baro-diffusion are predominant
for large Lewis numbers. Both coefficients are plotted in Fig. \ref{fig:Coeff Dc Dp}
for increasing $Z_{2}$. The former decreases with $Z_{2}$, while
the latter increases. Baro-diffusion is zero in both pure liner $\left(y=0\right)$
and fuel $\left(y=1\right)$ limits, and it dominates over concentration
gradient diffusion in most part of the layer when the liner atomic
number is large. As can be seen in Fig. \ref{fig:Coeff Dc Dp}(b),
the sum of both coefficients approaches $\mathcal{D}_{c}+\mathcal{D}_{p}=1-y$
when $Z_{2}$ is large. 

The system of normalized governing equations consists therefore of
Eqs. \eqref{eq:cont self-similar dif} and \eqref{eq:eq dif self-similar},
with density given by Eq. \eqref{eq:varrho}, plasma velocity by Eq.
\eqref{eq:v diff1} and fuel drift velocity by Eq. \eqref{eq:U1}.
It only depends on the free parameter $\text{Le}$, or, equivalently,
the liner atomic number. 

\section{results with mass diffusion\label{sec:results-with-mass-transport}}

\begin{figure}
\includegraphics[scale=0.25]{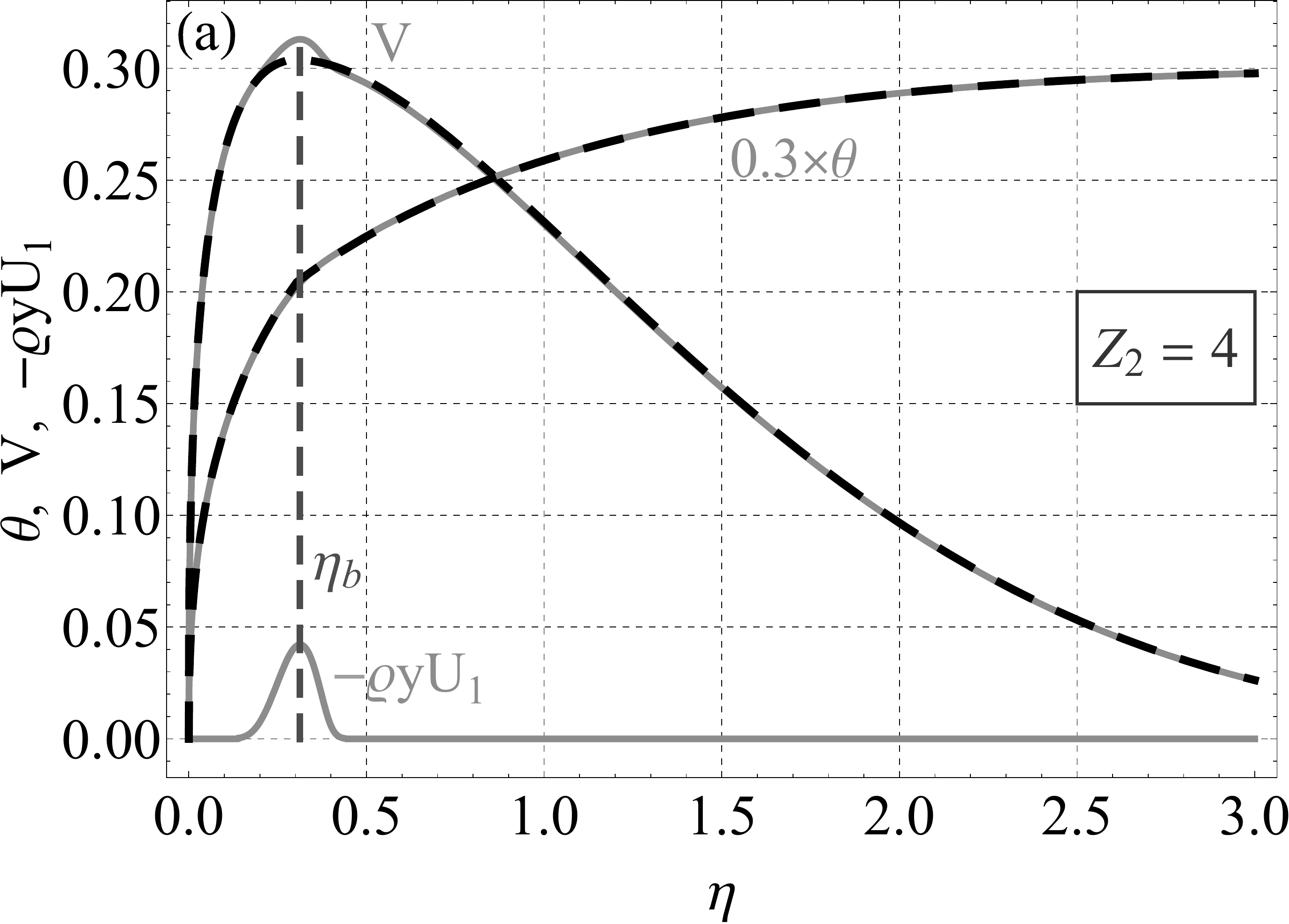}

\includegraphics[scale=0.25]{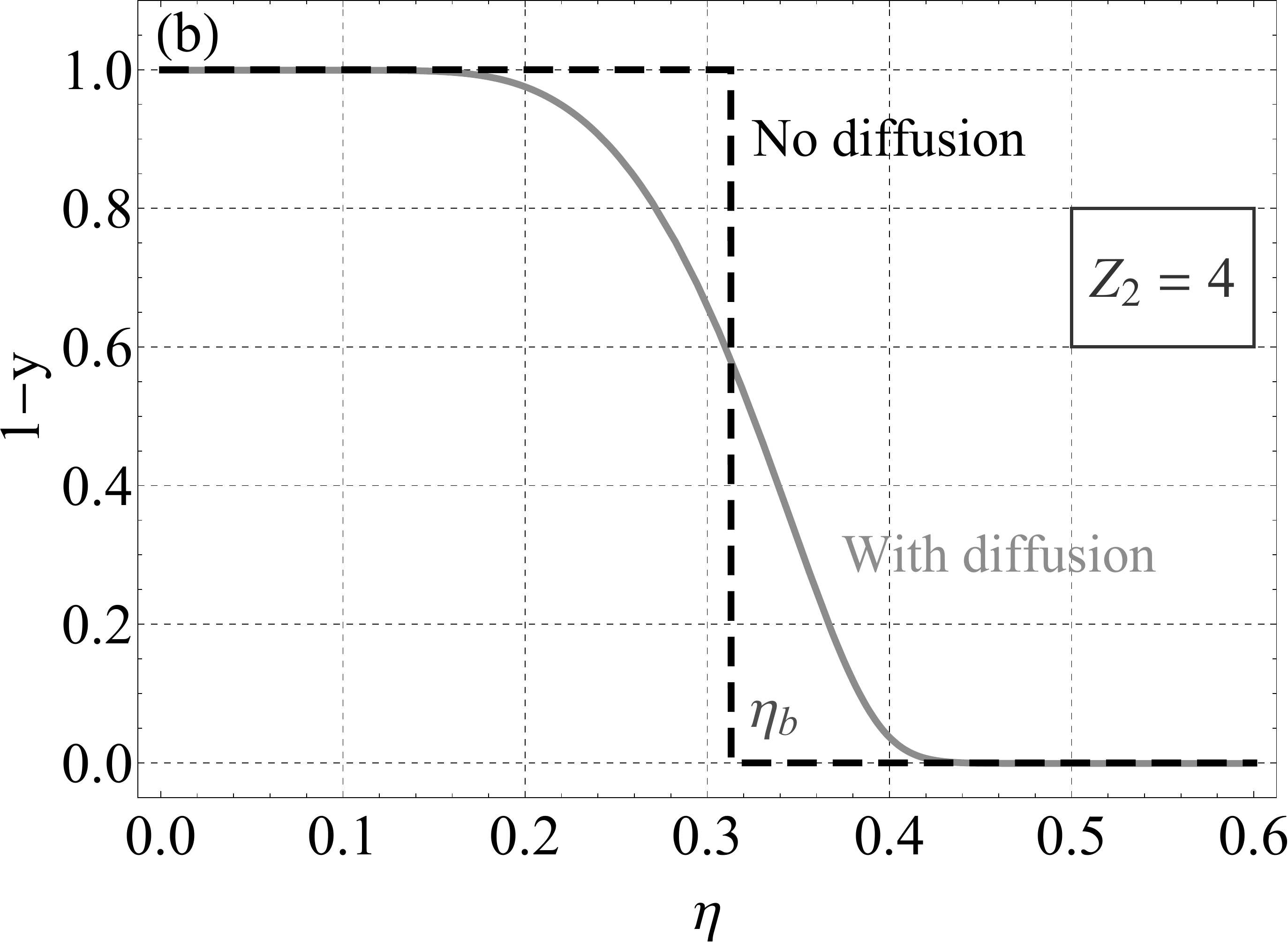}

\includegraphics[scale=0.25]{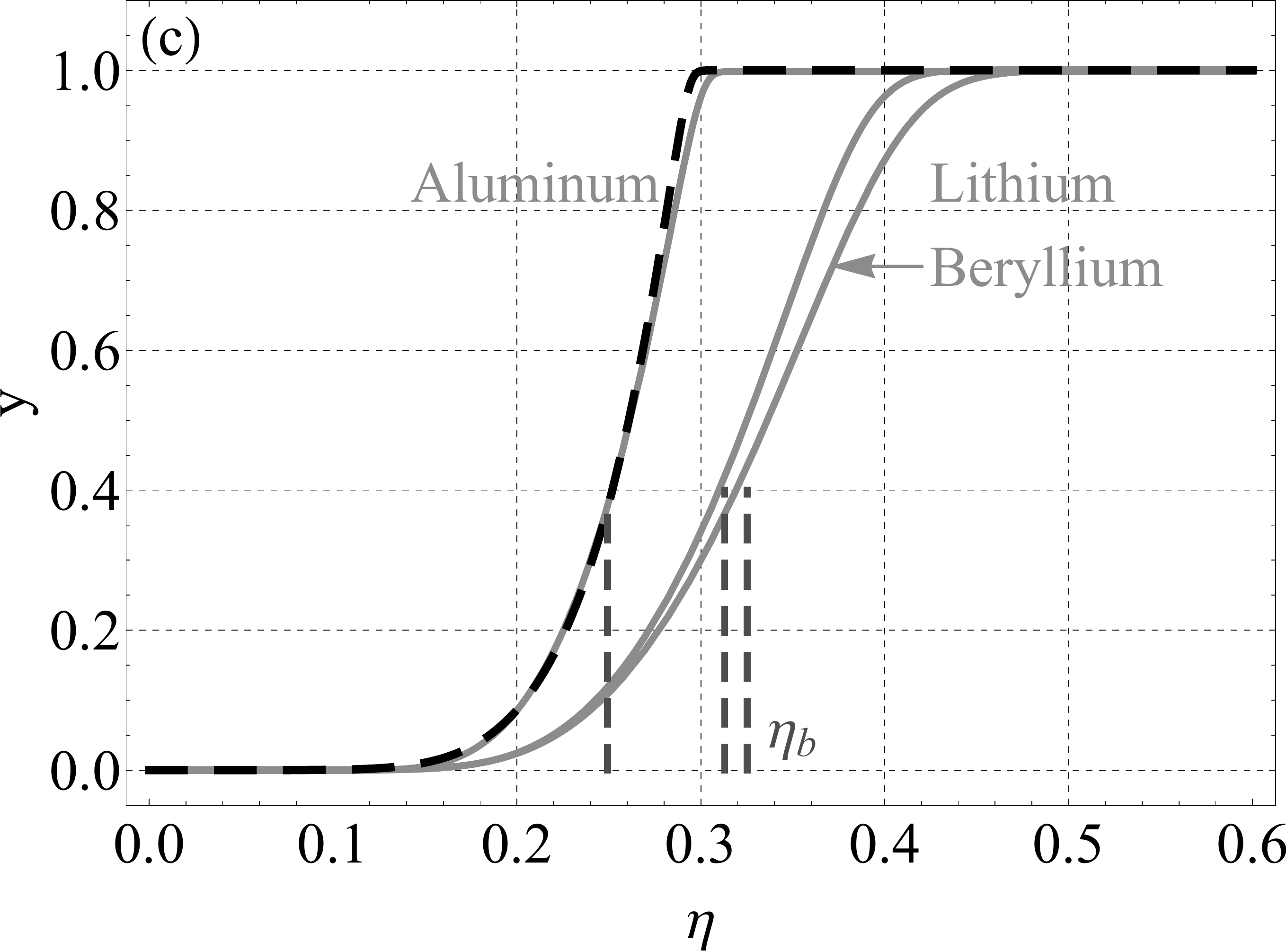}

\caption{\label{fig:Diffusion Profiles}(a) Normalized temperature $\theta$,
velocity $V$ and fuel mass diffusion rate $-\varrho yU_{1}$ for
a liner made of beryllium, depicted in solid gray lines. Temperature
and velocity profiles are compared to the solution without mass diffusion,
plotted in dashed black lines. (b) Beryllium mass concentration profile,
$\rho_{2}/\rho=1-y$, in solid gray lines compared to the solution
without mass diffusion, plotted in dashed lines. (c) Fuel mass concentration
$y$ for different liner materials. The black dashed line is the fuel
mass concentration profile for aluminum liner obtained from the boundary
layer model derived in Subsec. \ref{subsec:Mass-diffusion-boundary}.
The position of the ablated border $\eta_{b}$ is marked in gray dashed
lines. }
\end{figure}

The numerical resolution of the normalized governing equations is
shown in Fig. \ref{fig:Diffusion Profiles}. Temperature, velocity
and mass diffusion rate profiles are plotted in Fig. \ref{fig:Diffusion Profiles}(a)
for deuterium fuel and beryllium liner. Temperature and velocity profiles
are compared to the solution neglecting mass diffusion explained in
the first part of this paper. Both solutions should be identical for
an infinite Lewis number. It can be seen that, although the Lewis
number for beryllium is not significantly large, $\text{Le}=72$,
the profiles are remarkably similar. This suggests that, in a MagLIF
context, the hydrodynamic motion would not need to be solved self-consistently
with the mass diffusion problem. It can rather be solved independently
(immiscible plasmas), and mass diffusion be computed afterwards yielding
a similar result. This argument is reinforced by the results shown
in Table \ref{tab:Lewis}, where the position of the ablated border
and mass ablation are computed with and without mass diffusion, showing
good agreement. The fuel mass diffusion rate, $-\varrho yU_{1}$,
is relatively small and has a maximum at the ablated border. 

The most significant difference between the diffusion and no diffusion
solutions lies in the plasma velocity profile, which presents a bump
at the ablated border. It is there where liner and fuel materials
are in contact and mass diffusion takes place. As a consequence of
the strong fuel concentration variations, pressure inhomogeneities
arise, which locally accelerate the plasma. The plasma velocity is
modified and convects fuel material in order to restore isobaricity. 

Liner mass concentration profiles with and without mass diffusion
are plotted in Fig. \ref{fig:Diffusion Profiles}(b). Without diffusion,
the liner concentration drops from 1 to 0 at the ablated border. When
mass diffusion is taken into account, the ablated liner material penetrates
into the fuel a relatively short distance compared to the characteristic
thermal length, given by $\eta_{b}$. In Fig. \ref{fig:Diffusion Profiles}(c)
fuel mass concentration profiles are shown for different liner materials.
It can be seen that the width of the diffusion layer shrinks when
$Z_{2}$ increases, as the Lewis number becomes higher.

\begin{table}
\begin{tabular}{c|c|c|c}
Liner material & Lithium & Beryllium & Aluminum\tabularnewline
\hline 
\hline 
 &  &  & \tabularnewline
$\text{Le}$ & 85 & 111 & 323\tabularnewline
 &  &  & \tabularnewline
\hline 
 &  &  & \tabularnewline
$\left.\eta_{b}\right|_{\text{Diff.}}$ & 0.325 & 0.313 & 0.249\tabularnewline
 &  &  & \tabularnewline
\hline 
 &  &  & \tabularnewline
$\left.\eta_{b}\right|_{\text{No diff.}}$ & 0.315 & 0.304 & 0.245\tabularnewline
 &  &  & \tabularnewline
\hline 
 &  &  & \tabularnewline
$\left.\dfrac{m}{\rho_{0}\sqrt{\kappa_{0}t}}\right|_{\text{Diff.}}$ & 1.15 & 1.16 & 0.973\tabularnewline
 &  &  & \tabularnewline
\hline 
 &  &  & \tabularnewline
$\left.\dfrac{m}{\rho_{0}\sqrt{\kappa_{0}t}}\right|_{\text{No diff.}}$ & 1.12 & 1.13 & 0.951\tabularnewline
 &  &  & \tabularnewline
\hline 
 &  &  & \tabularnewline
$\dfrac{m_{lf}}{\rho_{0}\sqrt{\kappa_{0}t}}$ & 0.0471 & 0.0422 & 0.0272\tabularnewline
 &  &  & \tabularnewline
\hline 
 &  &  & \tabularnewline
$\left.\dfrac{m_{lf}}{\rho_{0}\sqrt{\kappa_{0}t}}\right|_{\text{B.L.}}$ & 0.0425 & 0.0388 & 0.0258\tabularnewline
 &  &  & \tabularnewline
\hline 
 &  &  & \tabularnewline
$100\times\dfrac{m_{lf}}{m}$ & 4.09\% & 3.64\% & 2.80\%\tabularnewline
 &  &  & \tabularnewline
\hline 
 &  &  & \tabularnewline
$\epsilon_{d}$ & 0.0753 & 0.0682 & 0.0471\tabularnewline
 &  &  & \tabularnewline
\hline 
 &  &  & \tabularnewline
$\dfrac{h_{\text{D}}}{gt^{2}}$ & 0.926 & 0.871 & 0.756\tabularnewline
 &  &  & \tabularnewline
\end{tabular}\caption{Numerical values for deuterium fuel, $Z_{1}=1$, and liners made of
lithium, $Z_{2}=3$, beryllium, $Z_{2}=4$, and aluminum, $Z_{2}=13$.\label{tab:Lewis}}
\end{table}

\subsection{Mass diffusion boundary layer\label{subsec:Mass-diffusion-boundary}}

When the Lewis number is large, mass diffusion is confined within
a sharp boundary layer placed at the ablated border. We recall that
the position of the ablated border is obtained by Eq. \eqref{eq:eta b}
and that velocity has null derivative there. 

In order to obtain the structure of the thin diffusive layer, we expand
the independent variable as $\eta=\eta_{b}+\epsilon_{d}s$, with $\epsilon_{d}\ll1$.
The fuel concentration varies from 0 to 1 in this region, hence we
assume $\text{d}y/\text{d}s\sim O\left(1\right)$. Since the layer
is thin, the temperature does not vary significantly in it, and we
make the isothermal approximation $\theta\approx\theta_{b}$, where
$\theta_{b}$ stands for the temperature value at the ablated border.
This hypothesis is consistent with the analysis of a diffusing gas-metal
interface in a thermonuclear plasma made in Ref. \onlinecite{molvig2014nonlinear},
where the authors assumed both isobaric and isothermal conditions.
Taking the first derivative in the equation of state \eqref{eq:varrho},
it can be obtained that $\text{d}\varrho/\text{d}s\sim\text{d}y/\text{d}s\sim O\left(1\right)$;
that is, density variation is important in this layer. According to
Eq. \eqref{eq:cont self-similar dif}, this variation forces the first
derivative of the velocity, $\text{d}V/\text{d}\eta$, to be of order
unity as soon as we move far from the ablated border while still being
inside the layer. Consequently, we expand $V\approx\eta_{b}+\epsilon_{d}V_{c}\left(s\right)$,
with $\text{d}V_{c}/\text{d}s\sim O\left(1\right)$. In the isothermal
approximation, and defining $\mathcal{D}_{y}\equiv\mathcal{D}_{c}+\mathcal{D}_{p}$,
the fuel drift velocity simplifies to 
\begin{equation}
U_{1}=-\dfrac{2\theta_{b}^{7/2}}{\text{Le}}\dfrac{\mathcal{D}_{y}}{y}\dfrac{\text{d}y}{\text{d}\eta}.\label{eq:U1 bl}
\end{equation}
Notice that only concentration gradients and baro-diffusion drive
mass mixing in the layer, while thermo-diffusion is negligible due
to the smallness of the temperature derivatives compared to the fuel
concentration derivatives. Inserting these expressions into the diffusion
equation \eqref{eq:eq dif self-similar}, and forcing the mass diffusion
term to be important, we can obtain the width of the diffusion layer
\begin{equation}
\epsilon_{d}=\sqrt{\dfrac{2\theta_{b}^{7/2}}{\text{Le}}}.\label{eq:width mass diffusion layer}
\end{equation}
The fuel drift velocity is therefore of order $O\left(1/\sqrt{\text{Le}}\right)$,
\begin{equation}
U_{1}=-\sqrt{\dfrac{2\theta_{b}^{7/2}}{\text{Le}}}\dfrac{\mathcal{D}_{y}}{y}\dfrac{\text{d}y}{\text{d}s}.\label{eq:U1 bl 2}
\end{equation}

Defining $a\equiv\mu\left(Z_{1}+1\right)/\left(Z_{2}+1\right)-1$,
the continuity and diffusion equations \eqref{eq:cont self-similar dif},
\eqref{eq:eq dif self-similar} take the form
\begin{equation}
\left(V_{c}-s\right)\dfrac{a}{1+ay}\dfrac{\text{d}y}{\text{d}s}=\dfrac{\text{d}V_{c}}{\text{d}s},\label{eq: cont layer}
\end{equation}
\begin{equation}
\left(V_{c}-s\right)\dfrac{1}{1+ay}\dfrac{\text{d}y}{\text{d}s}=\dfrac{\text{d}}{\text{d}s}\left(\dfrac{D_{y}}{1+ay}\dfrac{\text{d}y}{\text{d}s}\right).\label{eq:dif layer}
\end{equation}
As boundary conditions, we impose $y\left(s\rightarrow-\infty\right)=0$,
$y\left(s\rightarrow\infty\right)=1$, and $V_{c}\left(s=0\right)=0$.
These equations can be combined, yielding
\begin{equation}
\left[a\left(\dfrac{\mathcal{D}_{y}}{1+ay}\dfrac{\text{d}y}{\text{d}s}-\zeta\right)-s\right]\dfrac{1}{1+ay}\dfrac{\text{d}y}{\text{d}s}=\dfrac{\text{d}}{\text{d}s}\left(\dfrac{\mathcal{D}_{y}}{1+ay}\dfrac{\text{d}y}{\text{d}s}\right),\label{eq:combination}
\end{equation}
where 
\begin{equation}
\zeta=\left.\dfrac{\mathcal{D}_{y}}{1+ay}\dfrac{\text{d}y}{\text{d}s}\right|_{s=0}\label{eq:zeta}
\end{equation}
is an eigenvalue that has to be obtained self-consistently with $y\left(s\right)$
applying the former three boundary conditions. This eigenvalue is
related to the fuel mass diffusion rate at the ablation front,
\begin{equation}
\left.-\varrho yU_{1}\right|_{s=0}=\sqrt{\dfrac{2\theta_{b}^{3/2}}{\text{Le}}}\left(a+1\right)\zeta\label{eq:mass rate bl}
\end{equation}

Finally, the first order correction to velocity can be obtained \emph{a
posteriori} from Eq. \eqref{eq: cont layer} as 
\begin{equation}
V_{c}=a\left(\dfrac{\mathcal{D}_{y}}{1+ay}\dfrac{\text{d}y}{\text{d}s}-\zeta\right).\label{eq:Vc}
\end{equation}

\begin{figure}
\includegraphics[scale=0.25]{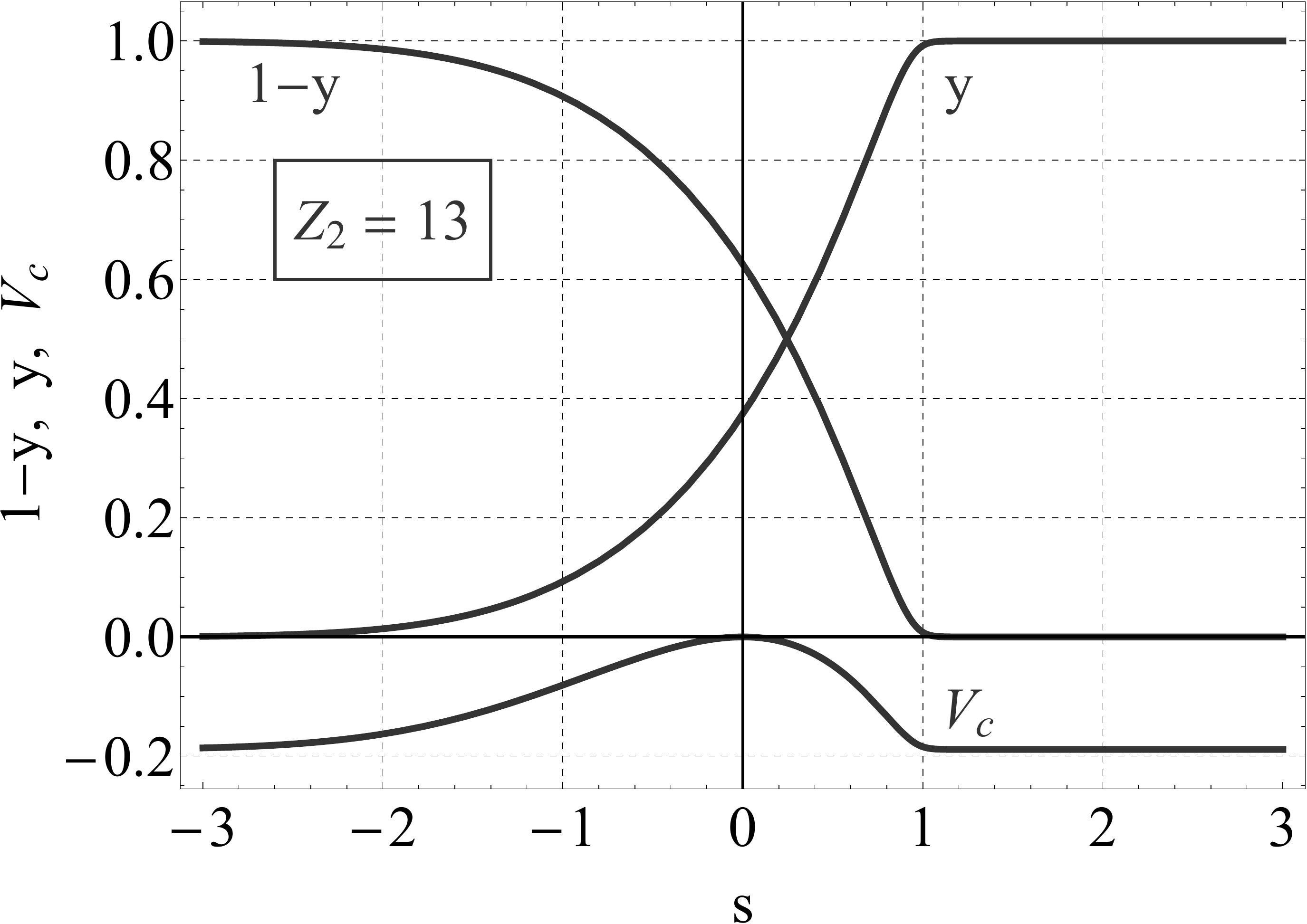}

\caption{\label{fig:boundary layers profiles}Profiles in the boundary layer
model. Fuel mass concentration, $y$, liner material mass concentration,
$1-y$, and first order correction for velocity, $V_{c}$, for aluminum
liner $\left(Z_{2}=13\right)$.}
\end{figure}

The structure of the layer for an aluminum liner is plotted in Fig.
\ref{fig:boundary layers profiles}. Since large Lewis number values
imply large $Z_{2}$, we use in this section the analytic expressions
\eqref{eq:norm coeff disparate mass} derived for a large $\mu$ ratio
for the transport coefficients. It can be seen that the profiles are
not symmetric. The asymmetry is introduced by the coefficient $\mathcal{D}_{y}$,
plotted in Fig. \ref{fig:Coeff Dc Dp}(b). The fuel mass concentration
profile presents a well defined wave front with clean fuel to the
right at $s\approx1$, and decreases exponentially to the left of
the ablated border, where the fuel diffuses into the ablated material.
Consequently, the parameter $\epsilon_{d}$ can be chosen as an accurate
characterization of the width of the diffusion layer. The first order
velocity correction $V_{c}$ corresponds to the bump appearing in
the velocity profile in Fig. \ref{fig:Diffusion Profiles}(a). The
eigenvalue $\zeta$ gives $\left\{ 0.25,0.24,0.22,0.21\right\} $
for $Z_{2}=\left\{ 3,4,13,\infty\right\} $, respectively. Although
not shown here, the profiles keep a similar structure for different
$Z_{2}$. Particularly, the wave front becomes a sharp corner when
$Z_{2}\rightarrow\infty$. The existence of a clear wave front was
also discussed in the analysis made in Ref. \onlinecite{molvig2014nonlinear}.
Comparisons of the boundary layer model with the complete solution
is shown in Fig. \ref{fig:Diffusion Profiles}(b), showing good agreement. 

The analysis of this layer presents slight differences compared to
the aforementioned study of a diffusing gas-metal interface in a thermonuclear
plasma, Ref. \onlinecite{molvig2014nonlinear}. Although the expressions
for the diffusion coefficients are the same (an equivalent notation
taken from Ref. \onlinecite{molvig2014classical} is used therein),
and the interface structure is also self-similar, the authors solved
the mass diffusion problem without taking into account the hydrodynamic
motion induced in the plasma. The governing equation of their diffusion
layer, Eq. (9) in Ref. \onlinecite{molvig2014nonlinear}, is equivalent
to Eq. \eqref{eq:dif layer} in this paper setting $V_{c}=0$. We
find that similar results are obtained when the hydrodynamic motion
of the plasma is not taken into account. Particularly, the eigenvalue
$\zeta$ is only modified by $1\%$.

\subsection{Liner material and fuel mixing}

We can characterize the fuel pollution by the amount of liner material
that diffuses into it, defined as
\begin{equation}
m_{lf}=\int_{x_{b}}^{\infty}\rho_{2}\text{d}x=\int_{x_{b}}^{\infty}\rho\left(1-y\right)\text{d}x.\label{eq:mass mixing}
\end{equation}
Taking the time derivative and making use of the diffusion equation
\eqref{eq:Difusion dif}, we obtain

\begin{equation}
\dfrac{\text{d}m_{lf}}{\text{d}t}=\left.-\rho yu_{1}\right|_{x_{b}}=-\dfrac{1}{2}\rho_{0}\sqrt{\dfrac{\kappa_{0}}{t}}\left.\varrho yU_{1}\right|_{\eta_{b}},\label{eq:mass mixing derivative}
\end{equation}
that is, fuel pollution is given by the mass diffusion rate at the
ablated border, where it is maximum. If we integrate this expression
in time, we have
\begin{equation}
\dfrac{m_{lf}}{\rho_{0}\sqrt{\kappa_{0}t}}=\left.-\varrho yU_{1}\right|_{\eta_{b}}.\label{eq:mass mixing result 1}
\end{equation}
We can use the boundary layer model, Eq. \eqref{eq:mass rate bl},
to further develop this expression in a large $\text{Le}$ limit,
\begin{equation}
\left.\dfrac{m_{lf}}{\rho_{0}\sqrt{\kappa_{0}t}}\right|_{\text{B.L.}}=\sqrt{\dfrac{2\theta_{b}^{3/2}}{\text{Le}}}\mu\dfrac{Z_{1}+1}{Z_{2}+1}\zeta.\label{eq:mass mixing result 2}
\end{equation}
The liner material diffused into the fuel scales therefore as the
inverse of the square root of the Lewis number. It is computed in
Table \ref{tab:Lewis} for different liner materials comparing the
exact result, Eq. \eqref{eq:mass mixing result 1}, with the boundary
layer model, \eqref{eq:mass mixing result 2}, giving similar results. 

We can also compute the fuel mass diffused into the ablated liner,
$m_{fl}=\int_{0}^{x_{b}}\rho_{1}\text{d}x=\int_{0}^{x_{b}}\rho y\text{d}x,$
which can be proved to be equal to the liner material diffused into
the fuel, $m_{fl}=m_{lf}$, by using Eqs. \eqref{eq:ion continuity dif}
and \eqref{eq:Difusion dif}. This is a consequence of choosing the
ablated border as the interface from which we compute fuel pollution.

As done in the analysis the problem without mass diffusion, we compute
the ablated liner material as
\begin{equation}
m\equiv\int_{0}^{\infty}\rho_{2}\text{d}x=\int_{0}^{\infty}\rho\left(1-y\right)\text{d}x.\label{eq:mass ablated with diff}
\end{equation}
Performing a similar manipulation, we can express this quantity as
\begin{equation}
\dfrac{m}{\rho_{0}\sqrt{\kappa_{0}t}}=\mu\dfrac{Z_{1}+1}{Z_{2}+1}\dfrac{\bar{K}_{e}\left(Z_{2}\right)}{\bar{K}_{e}\left(Z_{1}\right)}\dfrac{4}{5}s_{\theta}^{5/2},\label{eq:mass ablated normalized with diff}
\end{equation}
where $s_{\theta}$ is related to the temperature profile close to
the liner as defined in Eq. \eqref{eq:profiles close to the liner}.
Note that this expression for mass ablation is equivalent to its counterpart
in the problem without diffusion, Eq. \eqref{eq:mass normalized}. 

It is interesting to compute the percentage of mass ablation that
pollutes the liner, $100\times m_{lf}/m$, shown in Table \ref{tab:Lewis}.
It decreases from $4.09\%$ for lithium liner to $2.80\%$ for aluminum
liner. 

As commented in Sec. \ref{sec:analysis-and-results}, the steep negative
density gradient taking place at the ablated border may lead to a
Rayleigh-Taylor instability (RTI), causing liner and fuel mixing.
In order to compare fuel pollution caused by mass diffusion (microscopic
mixing) to the one due to the turbulent motion following the RTI (macroscopic
mixing), we roughly estimate the ratio $h_{\text{D}}/h_{\text{T}}$,
where $h_{\text{D}}=\epsilon_{d}\sqrt{\kappa_{0}t}$ is the width
of the diffusion layer, and $h_{\text{T}}\propto gt^{2}$ is the width
of the turbulent mixing layer. The constant of proportionality depends
on the particularities of the problem, and ranges from $0.03-0.07$\cite{youngs1984numerical,sanz2004nonlinear}.
We have then
\begin{equation}
\dfrac{h_{\text{D}}}{gt^{2}}=\dfrac{4\epsilon_{d}}{\eta_{b}},\label{eq:ratio layers}
\end{equation}
independently of time. This ratio is computed in Table \ref{tab:Lewis}.
It is of order unity and decreases slightly with $Z_{2}$, which indicates
that fuel pollution due to diffusion is as important as pollution
due to turbulent mixing. 

Finally, we can derive straightforward results in the limit $Z_{2}\rightarrow\infty$.
The Lewis number, given by Eq. \eqref{eq:Lewis}, can be taken as
$\text{Le}\approx23Z_{2}$. Since it is large, we can assume that
the hydrodynamic motion is uncoupled from the mass diffusion problem,
hence we take the results derived in Subsec. \ref{subsec:Analytic-solution-for}
for the position of the ablated border and mass ablation, and we assume
$\theta_{b}\approx1$. Consequently, we can derive 
\begin{equation}
\dfrac{m_{lf}}{\rho_{0}\sqrt{\kappa_{0}t}}\approx\dfrac{0.12}{\sqrt{Z_{2}}},\label{eq:mlf large Z2}
\end{equation}
\begin{equation}
\epsilon_{d}\approx\dfrac{0.29}{\sqrt{Z_{2}}}.\label{eq:ed large Z2}
\end{equation}
Which gives $100\times m/m_{lf}\approx2.36\%$ and $h_{\text{D}}/gt^{2}\approx0.72$,
independently of $Z_{2}$. 

\section{conclusions\label{sec:conclusions}}

The effect of the liner material on mass ablation, energy and magnetic
flux losses and liner - fuel mass diffusion have been studied in a
MagLIF fusion-like plasma. The self-similar evolution of a hot magnetized
fuel plasma in contact with a cold dense unmagnetized liner plasma
has been thoroughly described. 

In the first part of the paper, mass diffusion at the liner - fuel
interface (ablated border) has been neglected. The problem is governed
by heat conduction, and the fuel energy is lost in heating up the
ablated liner material. The magnetic field in the fuel is convected
by the Nernst velocity towards the liner. It penetrates into the ablated
liner and diffuses in a thin layer close to the ablation front. The
ablated border is a contact discontinuity, and the plasma density
at the liner side is greater than at the fuel side. This configuration
is Rayleigh-Taylor unstable as the ablated border penetrates into
the hot spot while being decelerated by the light fuel. The magnetic
field diffuses in a thin layer placed at the fuel side of the ablated
border and decreases its value to compensate the variation of the
Nernst velocity with the atomic number $Z$. The width of this diffusive
layer scales with the inverse of the magnetic Lewis number $\text{Le}_{\text{m}}$,
being therefore thinner than the diffusive layer at the ablation front,
which scales with the inverse of the square root of $\text{Le}_{\text{m}}$. 

For moderate and small magnetization levels, both thermal energy and
magnetic flux losses in the fuel decrease with the liner atomic number
$Z_{2}$, while mass ablation and magnetic flux losses in the hot
spot, composed by the ablated liner and fuel regions, present a maximum
for $Z_{2}=4$ and $Z_{2}=12$, respectively. An asymptotic analysis
performed in the large $Z_{2}$ limit shows that the four quantities
scale as $1/\sqrt{Z_{2}}$. 

In the second part of this paper, mass diffusion is taken into account,
but only the unmagnetized limit has been studied. The problem is governed
by the Lewis number $\text{Le}$, which only depends on the fuel and
liner atomic numbers and is an increasing function of the latter.
In a MagLIF context, it is typically large, hence heat conduction
governs the evolution of the problem, with mass diffusion being confined
within a thin layer placed at the ablated border. The width of this
layer scales with the inverse of the square root of $\text{Le}$.
Among the mechanisms giving rise to mass diffusion in a plasma, classical
diffusion due to concentration gradients is predominant for moderate
$Z_{2}$, while baro-diffusion becomes the most important for large
$Z_{2}$. Since the diffusion layer is thin and the temperature does
not change notably inside, thermo-diffusion results in a minor effect.
The temperature and velocity profiles are similar to the solution
without mass diffusion, and the mass diffusion problem can therefore
be solved \textit{a posteriori. }

The amount of liner material that diffuses into the fuel scales as
$1/\sqrt{Z_{2}}$. This mass represents a small percentage of the
total liner mass ablated into the hot spot. The percentage decreases
with $Z_{2}$ but attains a minimum of $2.36\%$ that cannot be further
reduced. Straightforward estimations suggest that the liner material
diffused pollutes the fuel in a layer that is comparable to the turbulent
mixing layer induced by the Rayleigh-Taylor instability that can develop
at the ablated border. This indicates that liner fuel mixing by microscopic
motion (diffusion) may therefore be as important as mixing due to
macroscopic motion (hydrodynamic instabilities leading to turbulence). 
\begin{acknowledgments}
This research was supported by the Spanish Ministerio de Economía
y Competitividad, Project No. ENE2014-54960R, and by the Spanish Ministerio
de Educación, Cultura y Deporte under the national research program
FPU, grant FPU 14/04879. We thank Professor Riccardo Betti from Laboratory
for Laser Energetics, University of Rochester, for fruitful discussion. 
\end{acknowledgments}

\appendix

\section{Review of transport coefficients\label{sec:Review-of-transport}}

In this Appendix, we summarize the theory already developed in Refs.
\onlinecite{simakov2014electron,molvig2014classical,simakov2016hydrodynamicI,simakov2016hydrodynamicII}
leading to the derivation of the electron and ion heat fluxes, $q_{e}$
and $q_{i}$, respectively, and the drift velocity $u_{1}$ in a two
ion species plasma. 

We first introduce the the electron-electron collision frequency 
\begin{equation}
\nu_{ee}=\dfrac{4\sqrt{2\pi}n_{e}e^{4}\log\Lambda_{ee}}{3m_{e}^{1/2}T_{e}^{3/2}},\label{eq:nuee}
\end{equation}
the electron-ion collision frequency 
\begin{equation}
\nu_{em}=\dfrac{4\sqrt{2\pi}n_{m}Z_{m}^{2}e^{4}\log\Lambda_{em}}{3m_{e}^{1/2}T_{e}^{3/2}},\label{eq:nuem}
\end{equation}
 and the ion-ion collision frequency 
\begin{equation}
\nu_{mn}=\dfrac{4\sqrt{\pi}n_{n}Z_{m}^{2}Z_{n}^{2}e^{4}\log\Lambda_{mn}}{3m_{m}^{1/2}T^{3/2}}.\label{eq:numn}
\end{equation}
Through all the appendix, the subscripts $m$ and $n$ only refer
to ion species. We assume the same value for the Coulomb logarithms
$\log\Lambda_{ee}=\log\Lambda_{em}=\log\Lambda_{mn}$.

The electron heat flux $q_{e}$ for a plasma with multiple ion species
was independently obtained in Refs. \onlinecite{simakov2014electron}
and \onlinecite{molvig2014classical} using a different variational
principle, but yielding the same result
\begin{equation}
q_{e}=-\dfrac{\gamma_{0}p_{e}}{m_{e}\left(\nu_{e1}+\nu_{e2}\right)}\nabla T_{e}.\label{eq:qe diff}
\end{equation}
 The coefficient $\gamma_{0}$ is a function of the effective ion
charge $Z_{\text{eff}}$, defined as
\begin{equation}
Z_{\text{eff}}\equiv\dfrac{\text{\ensuremath{\nu_{e1}}+\ensuremath{\nu_{e2}}}}{\nu_{ee}}=\dfrac{Z_{1}^{2}n_{1}+Z_{2}^{2}n_{2}}{Z_{1}n_{1}+Z_{2}n_{2}}\geq1,\label{eq:zeff}
\end{equation}
and reads

\begin{widetext}
\begin{equation}
\gamma_{0}\left(Z_{\text{eff}}\right)=\dfrac{25Z_{\text{eff}}\left(5,299,888Z_{\text{eff}}^{3}+21,559,755\sqrt{2}Z_{\text{eff}}^{2}+17,831,746Z_{\text{eff}}+1,272,672\sqrt{2}\right)}{4\left(2,447,104Z_{\text{eff}}^{4}+17,445,571\sqrt{2}Z_{\text{eff}}^{3}+57,670,090Z_{\text{eff}}^{2}+16,033,384\sqrt{2}Z_{\text{eff}}+2,013,696\right)}.\label{eq:gamma0}
\end{equation}
\end{widetext} The electron heat flux can be written into a Spitzer
form as
\begin{equation}
q_{e}=-\bar{K}_{e}\left(Z_{\text{eff}}\right)T^{5/2}\dfrac{\partial T}{\partial x},\label{eq:qe dif 2}
\end{equation}
with the Spitzer coefficient for the electron heat flux defined as
\begin{equation}
\bar{K}_{e}\left(Z_{\text{eff}}\right)\equiv\dfrac{3\gamma_{0}\left(Z_{\text{eff}}\right)}{4\sqrt{2\pi m_{e}}e^{4}\log\Lambda_{ee}Z_{\text{eff}}}.\label{eq:Ke}
\end{equation}
Note that this expression is the same as its counterpart for a single
ion species plasma, Eq. \eqref{eq:K no diff}, substituting $Z$ for
$Z_{\text{eff}}$.

Expressions for the ion drift velocities $u_{m}$ and the ion heat
flux $q_{i}$ have been derived in Ref. \onlinecite{simakov2016hydrodynamicI}
for an unmagnetized plasma with multiple ion species, and particularized
for a two ion species plasma in Ref. \onlinecite{simakov2016hydrodynamicII}.
As stated in these references, $u_{m}$ and $q_{i}$ are identified
as thermodynamic fluxes that can be related to their conjugate thermodynamic
forces $d_{m}$, $\partial\log T_{i}/\partial x$, respectively, through
a symmetric transport matrix
\begin{equation}
\left(\begin{array}{c}
u_{1}\\
\\
u_{2}\\
\\
q_{i}/p_{i}
\end{array}\right)=-\left(\begin{array}{ccc}
\Delta_{11} & \Delta_{12} & D_{T1}\\
\\
\Delta_{21} & \Delta_{22} & D_{T2}\\
\\
D_{T1} & D_{T2} & \kappa_{i}/n_{i}
\end{array}\right)\left(\begin{array}{c}
d_{1}\\
\\
d_{2}\\
\\
\dfrac{\partial\log T_{i}}{\partial x}
\end{array}\right),\label{eq:matrix coefficients or}
\end{equation}
with $\Delta_{12}=\Delta_{21}$ being the generalized diffusion coefficients,
$k_{i}$ being the ion heat conduction coefficient and $D_{Tm}$ being
the thermo-diffusion coefficients (the relation between $u_{m}$ and
$\partial\log T_{i}/\partial x$ is often referred to as the Ludwig-Soret
effect and reciprocal relation between $q_{i}$ and $d_{m}$ is referred
to as the Dufour effect). 

The thermodynamic force $d_{m}$ stands for the imbalance between
the fluid inertial force on the ion species $m$ ($\rho_{m}Dv/Dt$,
with $D/Dt$ being the substantial derivative) and the kinetic forces
acting on it. The diffusion fluxes arise to relax this imbalance,
the origin of which lies in the the collision between ion species.
It can be written as

\begin{multline}
d_{m}\equiv\dfrac{1}{p_{i}}\left\{ \dfrac{\partial p_{m}}{\partial x}-Z_{m}en_{m}E-F_{me}\right.\\
\left.-\dfrac{\rho_{m}}{\rho}\left(\dfrac{\partial p_{i}}{\partial x}-en_{e}E-\sum_{m}F_{me}\right)\right\} ,\label{eq:d1}
\end{multline}
with $E$ being the electric field and $F_{em}=-F_{me}=-\left(\beta_{0}n_{e}\nu_{em}/\sum_{m}\nu_{em}\right)\partial T_{e}/\partial x$
being the electron collisional friction force with the ion species
$m$, such that $\sum_{m}F_{em}=-\beta_{0}n_{e}\partial T_{e}/\partial x$.
The coefficient $\beta_{0}$ is a function of the effective ion charge,
$\beta_{0}\left(Z_{\text{eff}}\right)=30Z_{\text{eff}}\left(11Z_{\text{eff}}+15\sqrt{2}\right)/\left(217Z_{\text{eff}}^{2}+604\sqrt{2}Z_{\text{eff}}+288\right)$. 

The electric field can be obtained from the electron momentum equation
which, after neglecting electrons inertia and making use of the ambipolarity
condition, reads 
\begin{equation}
en_{e}E=-\dfrac{\partial p_{e}}{\partial x}-\beta_{0}n_{e}\dfrac{\partial T_{e}}{\partial x}.\label{eq:Efield}
\end{equation}
Inserting these relations into Eq. \eqref{eq:d1}, and taking into
account that $p_{m}=x_{m}p_{i}$, we can obtain more convenient expressions
for $d_{m}$

\begin{multline}
d_{m}=\dfrac{\partial x_{m}}{\partial x}+\left(x_{m}-\dfrac{\rho_{m}}{\rho}\right)\dfrac{\partial\log p_{i}}{\partial x}\\
+\left(\dfrac{\rho_{m}}{\rho}-\dfrac{Z_{m}n_{m}}{n_{e}}\right)\dfrac{n_{e}}{n}\dfrac{eE}{T_{i}}\\
+\left(y-\dfrac{Z_{m}^{2}n_{m}}{Z_{1}^{2}n_{1}+Z_{2}^{2}n_{2}}\right)\dfrac{n_{e}}{n}\dfrac{\beta_{0}}{T_{i}}\dfrac{\partial T_{e}}{\partial x},\label{eq:d1 expression 2}
\end{multline}
or, equivalently, using Eq. \eqref{eq:Efield}, 
\begin{multline}
d_{m}=\dfrac{\partial x_{m}}{\partial x}+\left(x_{m}-\dfrac{\rho_{m}}{\rho}\right)\dfrac{\partial\log p_{i}}{\partial x}\\
+\left(\dfrac{Z_{m}n_{m}}{n_{e}}-\dfrac{\rho_{m}}{\rho}\right)\dfrac{1}{p_{i}}\dfrac{\partial p_{e}}{\partial x}\\
+\left(\dfrac{Z_{m}n_{m}}{n_{e}}-\dfrac{Z_{m}^{2}n_{m}}{Z_{1}^{2}n_{1}+Z_{2}^{2}n_{2}}\right)\dfrac{n_{e}}{n}\dfrac{\beta_{0}}{T_{i}}\dfrac{\partial T_{e}}{\partial x}.\label{eq:d1 expression 3}
\end{multline}
Notice that, although the original expression for $d_{m}$, Eq. \eqref{eq:d1},
is the same as Eq. (20) in Ref. \onlinecite{simakov2016hydrodynamicI},
the more convenient expression \eqref{eq:d1 expression 3} differs
from its counterpart in the same reference, Eq. (25) therein, in the
coefficient $n_{e}/n$ multiplying the temperature derivative. We
suspect that this factor has been dropped by mistake, since Eq. \eqref{eq:d1 expression 3}
agrees with the definition of the same thermodynamic force in Eq.
(125) in Ref. \onlinecite{molvig2014classical}. 

The system \eqref{eq:matrix coefficients or} can be reduced noticing
that $\sum_{m}d_{m}=0$, yielding

\begin{equation}
\left(\begin{array}{c}
u_{1}\\
\\
q_{i}/p_{i}
\end{array}\right)=-\left(\begin{array}{cc}
\Delta_{11}^{\prime} & D_{T1}\\
\\
D_{T1}^{\prime} & \kappa_{i}/n_{i}
\end{array}\right)\left(\begin{array}{c}
d_{1}\\
\\
\dfrac{\partial\log T_{i}}{\partial x}
\end{array}\right),\label{eq:matrix coefficients}
\end{equation}
where $\Delta_{11}^{\prime}\equiv\Delta_{11}-\Delta_{12}$, $D_{T1}^{\prime}\equiv D_{T1}-D_{T2}$.
As done in Ref. \onlinecite{simakov2016hydrodynamicII}, it is convenient
to normalized the coefficients in Eq. \eqref{eq:matrix coefficients}
as
\begin{equation}
\begin{array}{cc}
\Delta_{11}^{\prime}=\dfrac{2T}{m_{1}\nu_{11}}\hat{\Delta}_{11}^{\prime}, & D_{T1}=\dfrac{2T}{m_{1}\nu_{11}}\hat{D}_{T1},\\
\\
D_{T1}^{\prime}=\dfrac{2T}{m_{1}\nu_{11}}\hat{D}_{T1}^{\prime}, & \kappa_{i}=\dfrac{2n_{1}T}{m_{1}\nu_{11}}\hat{\kappa}_{i}.
\end{array}\label{eq:normalized coeff}
\end{equation}
The fuel drift velocity can then be written as 
\begin{equation}
u_{1}=-\dfrac{2T}{m_{1}\nu_{11}}\hat{\Delta}_{11}^{\prime}\left(d_{1}+\dfrac{\hat{D}_{T1}^{\prime}}{\hat{\Delta}_{11}^{\prime}}\dfrac{\partial\log T_{i}}{\partial x}\right),\label{eq:u1}
\end{equation}
and we use this relation to express the ion heat flux in a more convenient
way
\begin{equation}
q_{i}=-\dfrac{2p_{i}T}{m_{1}\nu_{11}}\left(x_{1}\hat{\kappa}_{i}-\dfrac{\hat{D}_{T1}}{\hat{\Delta}_{11}^{\prime}}\hat{D}_{T1}^{\prime}\right)\dfrac{\partial\log T_{i}}{\partial x}+\dfrac{\hat{D}_{T1}^{\prime}}{\hat{\Delta}_{11}^{\prime}x_{1}}p_{1}u_{1},\label{eq:qi}
\end{equation}
which can therefore be written as a Spitzer term plus the contribution
due to mass diffusion
\begin{equation}
q_{i}=-\bar{K}_{i}T^{5/2}\dfrac{\partial T}{\partial x}+\dfrac{\hat{D}_{T1}^{\prime}}{\hat{\Delta}_{11}^{\prime}x_{1}}p_{1}u_{1}.\label{eq:qi bis}
\end{equation}
The Spitzer coefficient for the ion heat flux is 
\begin{equation}
\bar{K}_{i}\left(\upsilon\right)\equiv\dfrac{3}{2\sqrt{\pi}Z_{1}^{4}e^{4}\log\Lambda_{11}\sqrt{m_{1}}}\left(\hat{\kappa}_{i}-\dfrac{\hat{D}_{T1}\hat{D}_{T1}^{\prime}}{x_{1}\hat{\Delta}_{11}^{\prime}}\right).\label{eq:Ki}
\end{equation}
A straightforward estimation of the electron and ion conduction coefficients
gives $\bar{K}_{i}/\bar{K}_{e}\sim\sqrt{m_{e}/m_{p}}\ll1$, which
implies that $q_{i}$ is smaller than $q_{e}$, as happens in an unmagnetized
single ion species plasma. Nevertheless, ion heat flux is retained
in this analysis. 

In Eq. \eqref{eq:u1}, every mechanism contributing to ion diffusion
in an unmagnetized plasma can be identified. The first term of $d_{1}$,
Eq. \eqref{eq:d1 expression 2}, represents diffusion due to concentration
gradients, the second one stands for ion baro-diffusion, the third
one is the electro-diffusion, and the fourth one represents thermo-diffusion
due to electron temperature gradients. The Ludwig-Soret effect in
Eq. \eqref{eq:u1} stands for thermo-diffusion driven by ion temperature
gradients. Noticing that the fuel number density fraction $x_{1}$
can be related to the fuel mass concentration $y$ through
\begin{equation}
x_{1}=\dfrac{y}{y+\dfrac{1-y}{\mu}},\label{eq:x1 vs y}
\end{equation}
we can obtain the same baro and electro-diffusion ratios, $k_{p}$
and $k_{E}$, as derived by Kagan and Tang\cite{kagan2014thermodynamic}
in Eqs. (26) and (33) therein, without need to specify the transport
coefficients of Eq. \eqref{eq:matrix coefficients}. This confirms
the statement made in the same reference that the baro and electro-diffusion
ratios can be calculated uniquely by thermodynamic means, independently
of the nature of collisions. Expressing $d_{m}$ by Eq. \eqref{eq:d1 expression 2}
would correspond to choosing the ion mixture as the thermodynamic
system, hence only the ion pressure appears and the electric field
has to be taken into account. However, using Eq. \eqref{eq:d1 expression 3}
corresponds to choosing the plasma as a whole, then both ion and electron
pressure take place, and the electric field does not appear explicitly
as the plasma is quasi-neutral. Hereinafter, Eq. \eqref{eq:d1 expression 3}
will be used for $d_{1}$, since all the terms can be directly related
to temperature $T$ and fuel mass concentration $y$. 

The evaluation of the normalized coefficients in Eq. \eqref{eq:normalized coeff}
requires to solve a linear system of 6 equations for every value of
the fuel concentration $y$, as explained in Sec. III in Ref. \onlinecite{simakov2016hydrodynamicII}.
However, analytical solutions were derived in the same reference for
a mixture of ion species with disparate masses, $\mu\gg1$. Defining
$\zeta\equiv Z_{2}^{2}\log\Lambda_{12}/Z_{1}^{2}\log\Lambda_{11}=Z_{2}^{2}/Z_{1}^{2}$,
$\upsilon\equiv\nu_{12}/\nu_{11}=\left(n_{2}/n_{1}\right)\zeta=(1-y)\zeta/\mu y$,
they read
\begin{eqnarray}
 &  & \hat{\Delta}_{11}^{\prime}\left(y\right)=\dfrac{\mu\left(\zeta+\upsilon\right)}{\zeta\left(\zeta+\upsilon\mu\right)}\dfrac{217\sqrt{2}\upsilon^{2}+1.208\upsilon+288\sqrt{2}}{16\left(16\upsilon^{2}+61\sqrt{2}\upsilon+72\right)},\nonumber \\
 &  & \hat{D}_{T1}\left(y\right)=\dfrac{\upsilon\mu}{\zeta+\upsilon\mu}\dfrac{15\left(11\sqrt{2}\upsilon+30\right)}{8\left(16\upsilon^{2}+61\sqrt{2}\upsilon+72\right)},\nonumber \\
 &  & \hat{D}_{T1}^{\prime}\left(y\right)=\dfrac{15\left(11\sqrt{2}\upsilon+30\right)}{8\left(16\upsilon^{2}+61\sqrt{2}\upsilon+72\right)},\nonumber \\
 &  & \hat{\kappa}_{i}\left(y\right)=\dfrac{25\left(26\sqrt{2}\upsilon+45\right)}{8\left(16\upsilon^{2}+61\sqrt{2}\upsilon+72\right)}+\nonumber \\
 &  & \qquad\dfrac{125\upsilon\left(35\sqrt{2}+9\sqrt{\mu}\upsilon\right)}{4\zeta^{2}\left(5.250+1.375\sqrt{2\mu}\upsilon+144\mu\upsilon^{2}\right)}.\label{eq:norm coeff disparate mass}
\end{eqnarray}
In this paper, we use the evaluation of the coefficients for any $\mu$
rather than the analytical expressions derived for large $\mu$, with
the exception of the boundary layer model in Subsec. \ref{subsec:Mass-diffusion-boundary}.
Finally, we solved the linear system in the particular case $y=0$
to obtain $\hat{\Delta}_{11}^{\prime}\left(y=0\right)$ for any $\mu$,
required in the definition of the Lewis number, Eq. \eqref{eq:Lewis}.
It gives $\hat{\Delta}_{11}^{\prime}\left(0\right)=P_{\mu}/\zeta$,
with

\begin{widetext}
\begin{equation}
P_{\mu}=\sqrt{\dfrac{\mu+1}{\mu}}\dfrac{434\mu^{6}+3,912\mu^{5}+17,071\mu^{4}+33,152\mu^{3}+46,764\mu^{2}+38,080\mu+21,000}{16\sqrt{2}\left(16\mu^{6}+192\mu^{5}+1,064\mu^{4}+3,136\mu^{3}+5,058\mu^{2}+4,760\mu+2,625\right)}.\label{eq:Pmu}
\end{equation}
\end{widetext} 

\bibliographystyle{apsrev4-1}
\bibliography{references}

\end{document}